\documentclass[acmtog,screen]{acmart}

\usepackage{algorithm}
\usepackage{algpseudocode}
\usepackage{subcaption}
\usepackage{multirow}
\usepackage[table]{xcolor}
\usepackage{wrapfig}
\usepackage{graphicx}
\usepackage{float}

\newcommand{\x}{{\mathbf x}}
\newcommand{\V}{{\mathbf v}}
\newcommand{\hatx}{{\hat{\mathbf x}}}
\newcommand{\tildex}{{\tilde{\mathbf x}}}

\AtBeginDocument{%
  }

\setcopyright{acmlicensed}
\copyrightyear{2026}
\acmYear{2026}
\acmDOI{XXXXXXX.XXXXXXX}
\acmISBN{978-1-4503-XXXX-X/2026/06}

\begin{document}

\title{Robust and Efficient Penetration-Free Elastodynamics without Barriers}


\author{Juntian Zheng}
\email{juntianz@andrew.cmu.edu}
\affiliation{%
  \institution{Carnegie Mellon University}
  \country{USA}
}

\author{Zhaofeng Luo}
\email{zhaofen2@andrew.cmu.edu}
\affiliation{%
  \institution{Carnegie Mellon University}
  \country{USA}
}

\author{Minchen Li}
\email{minchernl@gmail.com}
\affiliation{%
  \institution{Carnegie Mellon University}
  \country{USA}
}
\affiliation{%
  \institution{Genesis AI}
  \country{USA}
}







\renewcommand{\shortauthors}{Juntian Zheng, Zhaofeng Luo, Minchen Li}

\begin{abstract}
We introduce a barrier-free optimization framework for non-penetration elastodynamic simulation that matches the robustness of Incremental Potential Contact (IPC) while overcoming its two primary efficiency bottlenecks: (1) reliance on logarithmic barrier functions to enforce non-penetration constraints, which leads to ill-conditioned systems and significantly slows down the convergence of iterative linear solvers; and (2) the time-of-impact (TOI) locking issue, which restricts active-set exploration in collision-intensive scenes and requires a large number of Newton iterations.
We propose a novel second-order constrained optimization framework featuring a custom augmented Lagrangian solver that avoids TOI locking by immediately incorporating all requisite contact pairs detected via CCD, enabling more efficient active-set exploration and leading to significantly fewer Newton iterations. By adaptively updating Lagrange multipliers rather than increasing penalty stiffness, our method prevents stagnation at zero TOI while maintaining a well-conditioned system. We further introduce a constraint filtering and decay mechanism to keep the active set compact and stable. 
A comprehensive set of experiments demonstrates the efficiency, robustness, finite-step termination, and first-order time integration accuracy of our method under a cumulative TOI-based termination criterion.
With a GPU-optimized simulator design, our method achieves an up to 103$\times$ speedup over GIPC on challenging, contact-rich benchmarks -- scenarios that were previously tractable only with barrier-based methods.
Our code and data are open-sourced at \url{https://simulation-intelligence.github.io/barrier-free}.
\end{abstract}


\begin{teaserfigure}
  \includegraphics[width=\textwidth]{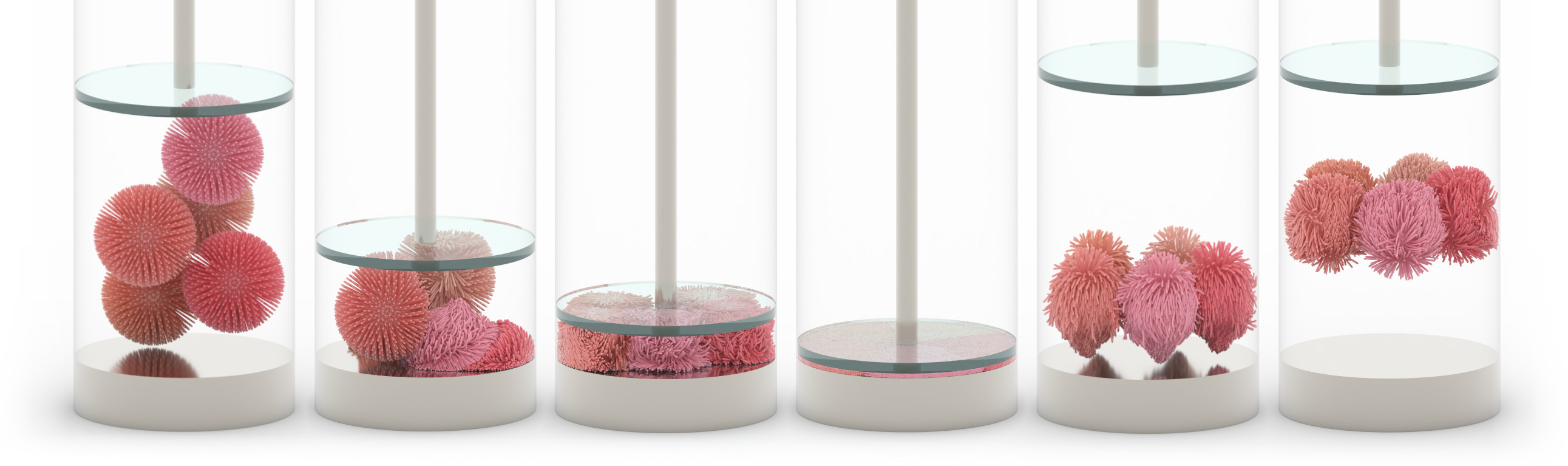}
  \caption{\textbf{Squishy balls under extreme compression.} Five elastic squishy balls are compressed by a moving boundary to extreme stress, generating dense contacts, and then released to rebound. The scene contains 2.61M DoFs, 2.25M tetrahedra, and generates up to 1.45M active contact constraints. With significantly fewer Newton iterations and better conditioning, we achieve a 98.5$\times$ speedup over GIPC \citep{huang2024gipc}, averaging 5.37 s per frame.}
  \label{fig:teaser}
\end{teaserfigure}


\maketitle


\begin{figure*}[!t]
  \includegraphics[width=\linewidth]{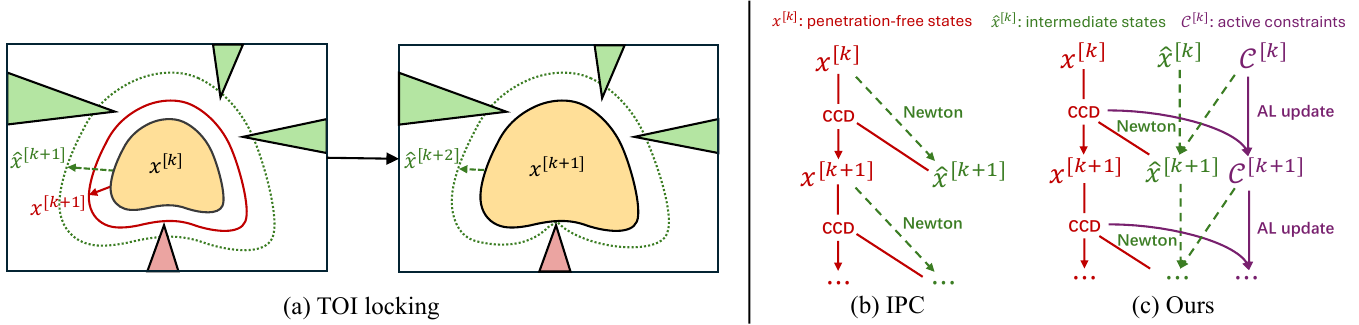}
  \caption{\textbf{Tackling the TOI locking issue.} (a) The advancement in each Newton iteration is stalled at the contact pair with the minimum TOI (marked in red), thus incorporating only the earliest contacts into the constraint set (assuming small contact radius). All other contacts (marked in green) can continue to block the CCD in subsequent iterations. (b) IPC's Newton iterations, discarding the solved intermediate states $\hat\x$. (c) Our modified framework with explicitly maintained intermediate state $\hat\x$ (initiated to $\mathbf{x}^t$ in each time step) and constraint set $\mathcal C$ carrying Lagrange multipliers, updated by a primal–dual augmented Lagrangian solver.}
  \label{fig:locking}
\end{figure*}

\section{INTRODUCTION}

In recent years, Incremental Potential Contact (IPC)~\cite{li2020incremental} has pioneered the penetration-free simulation of nonlinear elastic solids, offering guaranteed algorithmic convergence, solution accuracy, and minimal tuning of algorithmic parameters. IPC has been successfully applied to simulate a range of challenging phenomena. However, its computational efficiency remains a key bottleneck in time-sensitive applications such as robotics and virtual reality, even with recent GPU-accelerated variants featuring highly optimized Gauss–Newton and preconditioned conjugate gradient (PCG) solvers~\cite{huang2024gipc,huang2024stiffgipc}.

We identify two major sources of inefficiency in IPC: (1) the use of logarithmic barrier functions leads to severely \textit{ill-conditioned} systems, requiring many PCG iterations to solve; and (2) IPC suffers from the \emph{TOI locking} issue~\cite{lan2023second} in collision-intensive scenarios, where its filtered line search severely slows down active set exploration, a process that is inherently combinatorially complex in inequality-constrained optimization.

The \emph{TOI locking} problem arises because each Newton update is truncated by the smallest time-of-impact (TOI) detected via continuous collision detection (CCD)~\cite{wang2021large,li2020codimensional}, causing the earliest contact to stall the entire optimization step (\autoref{fig:locking}). As a result, many iterations are needed to progressively discover and incorporate all relevant contact pairs into the constraint set. Recent Gauss–Seidel-type strategies~\cite{lan2023second,chen2025offset} attempt to mitigate this issue using local updates, but remain limited in scenarios involving high stiffness or large deformations, due to their reliance on sublinearly convergent coordinate descent methods.

In this work, we propose a novel alternative that retains superlinearly convergent Newton iterations while \textit{improving active set exploration efficiency}. Our method immediately incorporates all requisite contacts detected by CCD into subsequent iterations, allowing earlier response to all these potential contacts. The key insight stems from reexamining IPC’s CCD-truncated Newton updates as shown in \autoref{fig:locking}: in each iteration, a possibly penetrating state $\hatx$ is generated from the previous penetration-free iterate $\x_\text{last}$, after which IPC applies CCD to obtain a new penetration-free state $\x$ by truncating the path between $\x_\text{last}$ and $\hatx$. IPC then discards $\hatx$ and proceeds from $\x$, potentially losing information of all contact pairs with larger TOI at $\hatx$. Instead, we resume Newton iterations directly from $\hatx$, allowing contacts with larger TOI to generate immediate responses.

This shift necessitates abandoning unsigned distances, whose gradients reverse upon penetration in $\hat\x$. We thus switch to using signed distances by linearizing the unsigned distance function at the last penetration-free iterate $\x_\text{last}$ during each update, which provides consistent and generalizable local contact force even in codimensional settings.
This immediately disables the use of the log-barrier, since it is undefined at a penetrating state $\hat{x}$ with negative distances. A seemingly straightforward alternative is to replace the barrier with a penalty-based collision response defined on the penetration depths in $\hatx$. However, constructing an effective penalty energy in this setting is nontrivial. Simple approaches such as a naïve quadratic penalty fail to guarantee that CCD-truncated Newton iterations will make consistent progress in challenging cases: even with a positive contact offset and large penalty stiffness, direct penalty methods can result in persistent penetrations across successive iterates of $\hatx$, causing CCD to repeatedly return zero TOI and thereby stalling progress (see \S\ref{sec:ablation}). To prevent such stagnation, a mechanism is required to strengthen contact response for persistent penetrations without indefinitely increasing the stiffness, which would otherwise degrade system conditioning just as the logarithmic barrier does.

To address this, we design a custom augmented Lagrangian (AL) solver~\cite{nocedal2006numerical}, which augments the penalty energy with iteratively updated Lagrange multipliers to provide improved control over constraint satisfaction. This enables persistent penetrations to be resolved by progressively adjusting the Lagrange multipliers without stiffening the penalty energy. As a result, the constraint violations converge toward a positive contact offset, enabling CCD to produce nonzero TOI and ensuring continued progress. This also yields \textit{better-conditioned systems} than the barrier-based approaches, significantly reducing the number of PCG iterations required even when using a simple block-Jacobi preconditioner (see \S\ref{sec:comparison}).

The use of linearized distance functions presents another key challenge: if the constraint set is not carefully managed, it may accumulate unnecessary and
potentially conflict constraints,
making the system overconstrained or even infeasible to solve \cite{li2020incremental}. Accordingly, we introduce a novel filtering scheme to avoid adding spatially irrelevant or redundant constraints and a decay mechanism to gradually phase out inactive ones. These techniques ensure that the constraint set remains compact and evolves smoothly across iterations, enabling efficient computation and preventing instability or oscillatory behavior in contact resolution.

To further improve efficiency without compromising simulation fidelity, we employ a termination criterion based on the cumulative TOI, extending heuristics used in~\cite{wang2023fast,ando2024cubic}. With high-level intuition and experimental evidence, we show that our method satisfies this criterion with tight tolerance in finite steps while achieving first-order time integration accuracy. In practice, our method consistently achieves larger average TOIs compared to IPC (see \S\ref{sec:ablation}), resulting in significantly fewer iterations to satisfy the same stopping condition. As a result, our method also tolerates looser termination thresholds without suffering from the damping artifacts commonly observed in IPC when limiting iteration counts.

Building upon these innovations, we present a novel elastodynamic simulator for penetration-free contact, achieving up to 103.15$\times$ speedup over GIPC~\cite{huang2024gipc} in challenging, contact-intensive benchmarks that were previously tractable only by barrier-based methods.
On moderate scenarios, our simulator (in double precision) achieves a $5.05\times$ speedup over Cubic Barrier~\cite{ando2024cubic} and a $33.1\times$ speedup over OGC~\cite{chen2025offset} using sufficient iterations to avoid artifacts, despite both baselines operating in single precision. In more challenging scenarios, our method outperforms Cubic Barrier by up to $84.4\times$, while OGC suffers from severe artifacts even when spending orders-of-magnitude more computation time than ours. These significant improvements are made possible by the following technical contributions:
\begin{itemize}
    \item A novel 2nd-order constrained optimization framework (\S\ref{sec:method-overview}), with a primal-dual augmented Lagrangian solver (\S\ref{sec:subproblem}) that ensures well-conditioned systems and consistent convergence progress, and a constraint filtering and decay mechanism (\S\ref{sec:activeset}) to achieve fast and robust active-set exploration.
    \item A GPU-optimized high-performance simulator design, featuring novel techniques for accelerated assembly of analytic elasticity Hessians under SPD projection (\S\ref{sec:gpu_opt}), conditioning-aware adjustment of penalty stiffness (\S\ref{sec:adaptive_mu}), and penalty-free enforcement of moving boundary conditions (\S\ref{sec:DBC}).
\end{itemize}


\section{RELATED WORK}

\subsection{Barrier-free Collision Handling}

We categorize all approaches not employing a diverging barrier function into the class of barrier-free methods. A straightforward way to handle collisions is through the \emph{penalty-based} methods \citep{benson1990single, wriggers1995finite, armero1998formulation, kim2020dynamic, chen2024vertex}, which introduce an energy term penalizing the penetration depth of geometric primitives into one another. Since the penalty energy is non-zero only when penetrations are present, the penalty-based methods inevitably allow penetrations in order to generate collision response. Even with a positive contact offset used to separate contacting primitives, the penalty stiffness must grow significantly to prevent penetration under high stress (see \S\ref{sec:ablation}).

Another traditional approach to collision handling, commonly used in cloth simulation, is the \emph{impact zone} method \citep{bridson2002robust, provot1997collision, tang2018cloth, li2020p, harmon2008robust, narain2012adaptive}, which groups penetrating primitive pairs into connected regions after solving the dynamics and projects each region back to a non-penetrating state. The projection is typically formulated as a linearly constrained optimization problem, solved using LCP solvers or first-order iterative methods. Despite their efficiency, impact zone methods lack the guarantee that all penetrations can be resolved within a finite number of iterations, which may lead to failure in highly complex contact configurations commonly encountered in elastodynamics simulations.

Fictitious domain methods (also known as air-mesh methods) offer an alternative approach that introduces a separate discretization of the void space and enforce global injectivity by maintaining non-negative volumes in both the material and air elements to handle contact \citep{muller2015air,jiang2017simplicial,misztal2012topology}. While this formulation provides a unified geometric constraint preventing inversion, it suffers from severe distortion-induced locking artifacts \citep{fang2021guaranteed}, necessitating frequent remeshing which can be highly inefficient in 3D.

Another class of collision handling approaches \citep{daviet2011hybrid, harmon2008robust, jean1992unilaterality, kane1999finite, kaufman2008staggered, kaufman2014adaptive, macklin2019non, otaduy2009implicit, verschoor2019efficient} is based on \emph{sequential quadratic programming (SQP)}, which formulates each time step as an optimization problem with non-penetration constraints, whose solution is approximated by sequentially solving a series of quadratic programming subproblems. The common pipeline of SQP-based methods involves, at each iteration, collecting the constraint set via discrete collision detection (DCD), forming a quadratic program with the quadratically approximated objective and linearized constraints, and solving it using LCP, conjugate residual (CR), Newton-like methods, or black-box QP solvers. The main limitations lie in the efficiency of solving the QP subproblems and the large number of iterations required to achieve a penetration-free state, which may even be impossible for complex scenarios under large time steps, as benchmarked in \citep{li2020incremental}.

It is noteworthy that some barrier-free methods are designed to generate penetration-free trajectories by applying strict collision detection to filter the advancing steps. Impact zone methods can be equipped with CCD to ensure that the iterations terminate upon reaching a penetration-free state \citep{tang2018cloth,li2020p}. \citet{wang2023fast} modified the CCD-based impact zone method by relaxing the linear penetration-free path constraint to a piecewise linear path and replacing CCD with a DCD-based displacement upper bound to ensure penetration-free iteration steps. \citet{lan2024efficient} introduced a PD framework with a CCD safeguard that dynamically adjusts contact constraint weights according to number of iterations they remain active. In challenging elastodynamic scenarios involving complex contact or high-speed impact, the iterations may stall in practice, requiring early termination for the simulation to proceed and potentially reducing fidelity. 
In comparison, our method is experimentally validated to terminate reliably under our TOI-based stopping criterion without sacrificing simulation fidelity, even in challenging scenarios.

\begin{figure*}[ht]
  \includegraphics[width=0.95\linewidth]{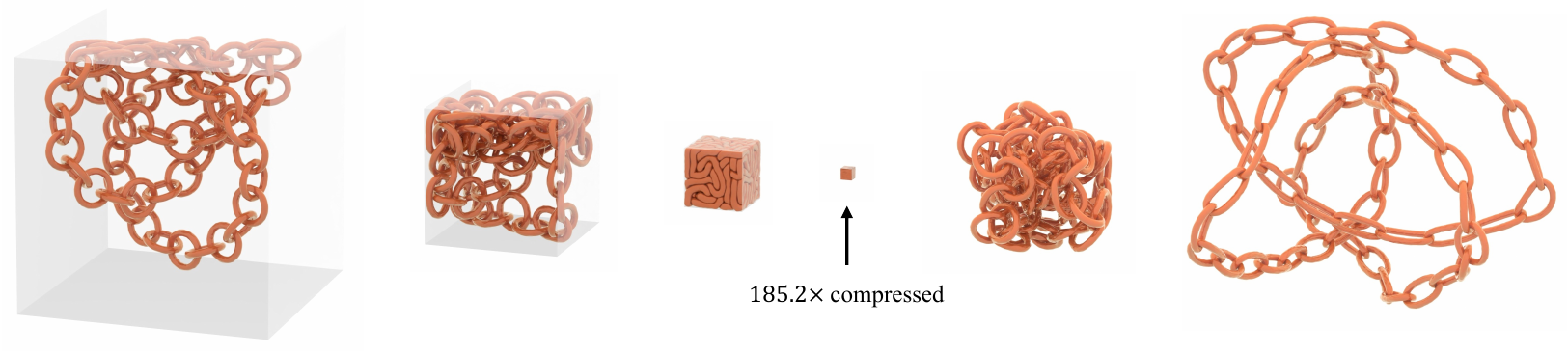}
  \caption{\textbf{Compressing chain rings.} Three nested elastic rings are compressed within a shrinking boundary and then released to rebound, reaching a density increase of up to $185.2\times$ during compression. The simulation remains stable and preserves topology under extreme deformation and complex contacts.}
  \label{fig:chains}
\end{figure*}

\subsection{Barrier-based Methods}

A more recent class of collision handling methods draws inspiration from the interior-point method, which introduces the logarithmic barrier function to ensure that the iterates remain within the feasible region (see \citep{nocedal2006numerical}). 
The pioneering work of \citet{li2020incremental} proposed Incremental Potential Contact (IPC), which achieves penetration-free large deformation elastodynamics through a $C^2$-continuous locally supported barrier function incorporated in the projected Newton framework with a CCD-truncated line search. A series of follow-up works have extended IPC to codimensional elements \citep{li2020codimensional}, rigid-body systems \citep{ferguson2021intersection,lan2022affine,chen2022unified}, granular impact dynamics \citep{jiang2022hybrid}, and coupling with MPM- \citep{li2024dynamic,li2022bfemp} and SPH-based \citep{xie2023contact} materials. \citet{li2023convergent} proposed a variational formulation of IPC's contact energies, identified discretization errors that lead to artificially distorted level sets near mesh edges and nodes, and introduced preliminary solutions to mitigate these issues.

Despite the improved accuracy and robustness in handling large deformations and complex contact scenarios, the main limitation of IPC lies in its low computational efficiency. A group of works have aimed to enhance its simulation efficiency by introducing various geometry representation, including medial elastics \citep{lan2021medial}, subspace DoFs \citep{du2025embedded, trusty2024trading}, and higher-order finite elements \citep{ferguson2023high,meng2025b}, to reduce the system DoFs. Another line of research focuses on improving the parallelization of IPC on modern GPUs. \citet{huang2024gipc} proposed GIPC, a fully GPU-optimized IPC framework with a parallelization-friendly Gauss-Newton approximation of the IPC barrier. \citet{du2024intersection} presented another GPU-parallelized IPC framework supporting efficient rigid-soft contact for robot manipulation. More recently, \citet{huang2024stiffgipc} further accelerated IPC for stiff materials by improving the Multilevel Additive Schwarz (MAS) \citep{wu2022gpu} linear solver preconditioner through connectivity enhancement and introducing a faster Hessian matrix assembly strategy for the affine-deformable coupled system.

Besides the standard projected Newton, the IPC barrier is also compatible with various optimization methods, including preconditioned nonlinear conjugate gradient \citep{shen2024preconditioned} and projective dynamics \citep{lan2022penetration,li2023subspace}. \citet{lan2023second} proposed a stencil descent method for IPC contact, which performs parallelized local hybrid Gauss–Seidel and Jacobi updates, and ensures penetration-free results via local-global CCD filtering. \citet{chen2025offset} smoothly extended the IPC barrier using a quadratic function and corrected its behavior under large contact radius through an offset geometry-based distance calculation, followed by energy optimization via displacement-bound vertex block descent \cite{chen2024vertex} updates. These methods mostly rely on low-order convergent optimizers to trade accuracy of the momentum equation for efficiency, which can easily lead to artificially damped motions. In contrast, \citet{guo2024barrier} retains the superlinearly convergent Newton iterations and introduces Lagrangian terms for contact pairs with small distances to improve system conditioning. However, their formulation still relies on logarithmic barrier functions, and its efficiency remains sensitive to the tuning of multiple algorithmic parameters.

In addition to the logarithmic barrier, other types of barrier functions have been proposed to handle penetration-free contact, including the non-local tangent-point energy \citep{strzelecki2013tangent} used in geometry processing \citep{sassen2024repulsive}, and the recently proposed locally supported Geometric Contact Potential \citep{huang2025geometric} that avoids the spurious forces generated by the IPC barrier under large contact radius. \citet{ando2024cubic} introduces a cubic contact energy that adaptively adjusts the stiffness based on the elasticity Hessian and distance gap, exhibiting barrier-like behavior as the contact distance approaches zero.

\section{BACKGROUND}

We begin by introducing the essential background and notation appearing throughout our framework. We focus on simulating the dynamic behavior of a set of elastic bodies spatially discretized into linear tetrahedral elements. The system's state consists of the stacked nodal positions $\mathbf{x}\in\mathbb R^{3N}$ and velocities $\mathbf{v}\in\mathbb R^{3N}$, in which $N$ denotes the number of nodes after discretization. The continuous-time trajectories $\mathbf{x}(t),\mathbf{v}(t)$ are further discretized into timesteps $\mathbf x^t, \mathbf v^t$. In the implicit Euler time integration commonly used in physics-based animation, the discrete timesteps are updated by
\begin{equation} \label{eq:implicit-Euler}
\begin{split}
\mathbf x^{t+1}&=\mathbf x^t+h\mathbf v^{t+1},\\
\mathbf v^{t+1}&=\mathbf v^t+h\mathbf M^{-1}(\mathbf f_{\text {int}}(\mathbf x^{t+1})+\mathbf f_\text{ext}^{t+1}),
\end{split}
\end{equation}
in which $h$ denotes the timestep size, $\mathbf M$ denotes the lumped mass matrix, $\mathbf f_\text{int}$ and $\mathbf f_\text{ext}$ denote the total internal and external forces, respectively. Let $U(\mathbf x)$ be the total potential energy associated with the internal forces (i.e., $\mathbf{f}_\text{int}(\mathbf x)=-\partial U/\partial\mathbf{x}$), including the smoothed friction energy as in IPC (detailed in Appendix~\ref{sec:friction}). By introducing the Incremental Potential \cite{kane2000variational}
\begin{equation}
E(\mathbf x,\tildex)=\frac12(\mathbf x-\tilde{\mathbf x})^T\mathbf M(\mathbf x-\tilde{\mathbf x})+ h^2 U(\mathbf x),
\end{equation}
where $\tilde{\mathbf x}=\mathbf x^t+h\mathbf v^t+h^2\mathbf M^{-1}\mathbf f_\text{ext}^t$, we can reformulate the nonlinear system (\autoref{eq:implicit-Euler}) into an equivalent minimization problem over $E(\mathbf x,\tilde{\mathbf x})$ w.r.t. $\mathbf{x}$, followed by velocity updates.

When an intersection-free trajectory is required, positive distance constraints are imposed on every intermediate state during the transition between adjacent timesteps \cite{li2020incremental}, i.e., there exists a path $\mathbf L_{\x^t\to\x^{t+1}}$ connecting $\x^t$ and $\x^{t+1}$ such that
\begin{equation}
\mathbf d(\x)> 0,\forall \x\in \mathbf L_{\x^t\to\x^{t+1}},
\end{equation}
where $\mathbf d=\{d_i\}_{i\in \mathcal I}$ is the set of unsigned distances between all vertex-face and edge-edge pairs. Inspired by the interior-point method, \citet{li2020incremental} proposed IPC by adding smoothly clamped log barriers to the incremental potential to handle the constraints:
\begin{equation}
B(\x)=E(\x,\tildex)+\kappa\sum_{i\in\mathcal I}b(d_i(\x),\hat d),
\end{equation}
where $\kappa,\hat d$ are the barrier parameters, and $b(d,\hat d)$ is a barrier function supported on $d\in(0,\hat d)$ that diverges as $d\rightarrow 0$. An intersection-free trajectory is then ensured by iteratively minimizing quadratic proxy of $B(\x)$ using line search with projected Newton while clamping the search directions using CCD.

Despite its robustness, accuracy, and penetration-free guarantees, IPC exhibits significant efficiency bottlenecks. These primarily stem from the clamping of search directions and the severe ill-conditioning induced by the sharp logarithmic barrier functions. 

\section{METHOD} \label{sec:method}

\subsection{Overview}
\label{sec:method-overview}

\paragraph{Problem Formulation}
Instead of relying on the log-barrier function as in IPC, we explore an alternative approach to advance $\mathbf x^t$ to the next timestep along an intersection-free trajectory. Rather than using the proximity distances to generate the contact constraints, we explicitly maintain an active constraint set $\mathcal C$ including all primitive pairs potentially leading to intersections, enabling earlier contact response during the solve for more effective iterations. In each timestep, we divide the time-stepping into a sequence of subproblems with linear inequality constraints:
\begin{equation} \label{eq:sub-problem}
\hat{\mathbf x}^{[k]}=\text{arg}\min_{\hat{\mathbf x}} E(\hat{\mathbf x},\tildex)\quad\text{s.t.}\quad \mathbf c^{[k]}(\hat{\mathbf x})\ge 0,
\end{equation}
in which $\mathbf c^{[k]}$ is a set of linearized contact constraints generated from the current active set $\mathcal C^{[k]}$ (see \S \ref{sec:subproblem}). In each iteration $k$, we perform one or several Newton steps to obtain an inexact solution of $\hat x^{[k]}$ before proceeding to the next iteration. 

\paragraph{Constructing Penetration-Free Paths}
The active sets $\mathcal C^{[k]}$ are progressively updated along with the intermediate states $\hatx^{[k]}$, and may not contain all necessary constraints in the early iterations. Due to the incompleteness of $\mathcal C^{[k]}$ and the linearization of the constraints, $\{\hat{\mathbf x}^{[k]}\}_{k\ge 1}$ are \emph{not guaranteed to be intersection-free}. Based on the intermediate states, we further construct an intersection-free piecewise linear path $\{\mathbf x^{[k]}\}_{k\ge 0}$ starting from $\mathbf x^{[0]}=\mathbf x^t$, where each $\mathbf x^{[k+1]}$ is a linear interpolation between $\mathbf{x}^{[k]}$ and $\hat{\mathbf{x}}^{[k+1]}$:
\begin{equation} \label{eq:CCD-clamp}
\mathbf x^{[k+1]}=(1-\alpha^{[k+1]})\mathbf x^{[k]}+\alpha^{[k+1]}\hat{\mathbf x}^{[k+1]},\quad 0\le \alpha^{[k+1]}\le 1,
\end{equation}
such that no intersections occur during the linear transition from $\mathbf x^{[k]}$ to $\mathbf x^{{[k+1]}}$, which can be realized using CCD. 
This decoupling of intersection-free states and solver iterates resembles IPC's quadratic proxy minimization followed by filtered line search. However, rather than discarding the proxy solution, we leverage it to anticipate potential contact constraints and to guide the estimation of optimization progress, as detailed later.

\paragraph{Termination Criterion}
It turns out that each $\mathbf x^{[k]}$ is also a linear combination of the previous timestep $\mathbf x^t$ and the intermediate states $\{\hat{\mathbf x}^{[i]}\}_{i=1}^k$:
\begin{equation}
\mathbf x^{[k]}=\beta_0^{[k]}\mathbf x^t+\sum_{i=1}^k\beta_i^{[k]}\hat{\mathbf x}^{[i]}
\end{equation}
with coefficients
\begin{equation}
\beta_i^{[k]}=\alpha^{[i]}\prod_{j=i+1}^{k}(1-\alpha^{[j]}),\quad 0\le i\le k,
\end{equation}
where $\alpha^{[0]}=1$. 
When the step sizes $\alpha^{[k]}$ are sufficiently large, the influence of the initial state $\mathbf x^t$ and early intermediate states $\hat{\mathbf x}^{[1]}, \hat{\mathbf x}^{[2]}, \dots$ on $\mathbf x^{[k]}$ decays exponentially. Consequently, as we iteratively update the active sets $\mathcal C^{[k]}$ and the intermediate states $\hat{\mathbf x}^{[k]}$ while maintaining non-infinitesimal $\alpha^{[k]}$, we eventually reach an intersection-free state $\mathbf x^{[k]}$ that is primarily composed of later iterates. Since these later iterates are solved using more complete $\mathcal C^{[k]}$, they better capture the true contact interactions, making $\mathbf x^{[k]}$ more accurate. We terminate the solve by setting $\mathbf x^{t+1} = \mathbf x^{[k]}$ and computing $\mathbf v^{t+1}$ once the total weight of the first $K_\text{min}$ intermediate states falls below a small user-defined threshold $\epsilon$:
\begin{equation} \label{eq:termination}
\sum_{i=0}^{K_\text{min}-1}\beta_i^{[k]}<\epsilon.
\end{equation}
When $K_\text{min}=1$, \autoref{eq:termination} reduces to the stopping condition of \cite{wang2023fast,ando2024cubic}.
As we will demonstrate later, this criterion enables early termination while avoiding damping artifacts and maintaining reasonable accuracy for our method.

Algorithm~\ref{alg:time-stepping} outlines the main pipeline of our time-stepping solver. We use $\beta^{[k]}$ to denote $\sum_{i=0}^{K_\text{min}-1}\beta_i^{[k]}$, which remains equal to $1$ during the first $K_\text{min}-1$ iterations and is scaled by $1-\alpha^{[k]}$ in the subsequent iterations. Similar to IPC, \Call{MaxStepSize}{} computes a conservative step size $\alpha^{[k+1]}$ that ensures intersection-free (and inversion-free for non-invertible materials such as Neo-Hookean) trajectories. The two core subroutines, \Call{SolveSubproblem}{} and \Call{UpdateActiveSet}{}, are detailed in the following subsections.

\begin{algorithm}
\caption{Main pipeline of our time-stepping algorithm.}\label{alg:time-stepping}
\begin{algorithmic}[1]
\Require last timestep's state $\mathbf x^t,\mathbf v^t$ and active set $\mathcal C^t$.
\Ensure current timestep's state $\mathbf x^{t+1},\mathbf v^{t+1}$ and active set $\mathcal C^{t+1}$.
\State $\tilde {\mathbf x} \gets \mathbf x^t+h\mathbf v^t+h^2\mathbf M^{-1}\mathbf f_\text{ext}^t$;
\State $\mathbf x^{[0]},\hat{\mathbf x}^{[0]},\mathcal C^{[0]}\gets\mathbf x^t,\mathbf x^t,\mathcal C^t$;
\State Handle moving boundaries in $\hatx^{[0]}$;\Comment{See \S \ref{sec:DBC}}
\State $\beta^{[0]},k\gets1,0$;\Comment{$\beta^{[k]}$ stores $\sum_{i=0}^{K_\text{min}-1}\beta_i^{[k]}$}
\While{$\beta^{[k]}>\epsilon$}
\State $\hat {\mathbf x}^{[k+1]}\gets$ \Call{SolveSubproblem}{$\tilde{\mathbf x},\mathbf x^{[k]},\hat{\mathbf x}^{[k]},\mathcal C^{[k]}$};\\\Comment{Algorithm \ref{alg:subproblem}, \S \ref{sec:subproblem}}
\State $\mathcal C^{[k+1]}\gets$ \Call{UpdateActiveSet}{$\mathbf x^{[k]},\hat{\mathbf x}^{[k]},\mathcal C^{[k]}$};\\\Comment{Algorithm \ref{alg:activeset}, \S \ref{sec:activeset}}
\State $\alpha^{[k+1]}\gets$ \Call{MaxStepSize}{$\mathbf x^{[k]},\hat{\mathbf x}^{[k+1]}$};\Comment{CCD queries}
\State $\mathbf x^{[k+1]}\gets(1-\alpha^{[k+1]})\mathbf x^{[k]}+\alpha^{[k+1]}\hat{\mathbf x}^{[k+1]}$;
\If{$k+1\ge K_\text{min}$}
    \State $\beta^{[k+1]}\gets(1-\alpha^{[k+1]})\beta^{[k]}$;
\Else
    \State $\beta^{[k+1]}\gets\beta^{[k]}$;
\EndIf
\State $k\gets k+1$;
\EndWhile
\State \Return $\mathbf x^{[k]},(\mathbf x^{[k]}-\mathbf x^t)/h,\mathcal C^{[k]}$;
\end{algorithmic}
\end{algorithm}

Our framework is related to \citet{wang2023fast}, but differs in two key aspects. First, instead of splitting each update into an unconstrained solve followed by a geometric projection under linearized collision constraints, we solve the constrained subproblem using a unified augmented Lagrangian formulation. As the geometric projection is agnostic to the physical properties of the objects, decoupling the projection from the dynamics solve can introduce artifacts such as highly stressed regions, as shown in \autoref{fig:lcp}. Second, the backward projection in \citep{wang2023fast} is solved using projected Gauss–Seidel iterations, which are less expensive than Newton solvers but do not guarantee that all constraints are resolved, especially when the collision pattern becomes complex, as shown in the comparison in \autoref{sec:comparison}. Our method instead relies on an augmented Lagrangian mechanism to adaptively strengthen unsatisfied constraints, ultimately driving them to a non-penetrating state.


\paragraph{Challenges}
Several key challenges may arise within this framework.
First, efficiently solving the subproblem (\autoref{eq:sub-problem}) remains challenging. To ensure large TOI when using the solved search direction, the approximate solution of $\hat{x}^{[k]}$ needs to have small or even no constraint violations.
Second, the termination of this pipeline relies on the values of $\alpha^{[k]}$ being sufficiently large, which is directly related to the update strategy of the active set $\mathcal C^{[k]}$. A conservative strategy (e.g., not updating at all) can easily lead to getting stuck at $\alpha^{[k]}=0$ forever. Meanwhile, overly aggressive strategies may result in unnecessarily large active sets, thereby increasing computational cost and system instability (as the constraints are linearized). 
Finally, 
it is still questionable whether our TOI-based termination criterion is sufficiently strong, i.e., if it ensures the necessary order of convergence to the PDE solution as a time integrator.

To address these challenges, \S\ref{sec:subproblem} introduces an efficient yet effective solver for \autoref{eq:sub-problem} based on the Augmented Lagrangian formulation. Section~\ref{sec:activeset} presents our active set update strategy, and Section~\ref{sec:termination} discusses the highl-level intuition of our framework's finite step termination and first-order time integration accuracy.

\subsection{Augmented Lagrangian-based Subproblem Solver}
\label{sec:subproblem}

\paragraph{Constraint Linearization}
We first formulate the subproblem (\autoref{eq:sub-problem}) by defining the linear constraints $\mathbf c(\hatx)$. For each vertex-face or edge-edge pair indexed by $i$, the unsigned distance $d_i(\x)$ between two primitives can be linearized into a signed distance function by first-order Taylor expansion at an intersection-free state $\x$:
\begin{equation}
\mathbf c(\hatx)=\{c_i(\hatx):=d_i(\x)+\nabla d_i(\x)^T(\hatx-\x)-\delta\}_{i\in \mathcal I_\mathcal C},
\end{equation}
where $\mathcal I_\mathcal C$ denotes all primitive pairs in the current active set $\mathcal C$ and $\delta$ is the surface separation parameter similar to $\hat d$ in IPC.
Expanding at $\mathbf x$ ensures correct orientation of the linearized constraints.

\begin{figure}
  \includegraphics[width=\linewidth]{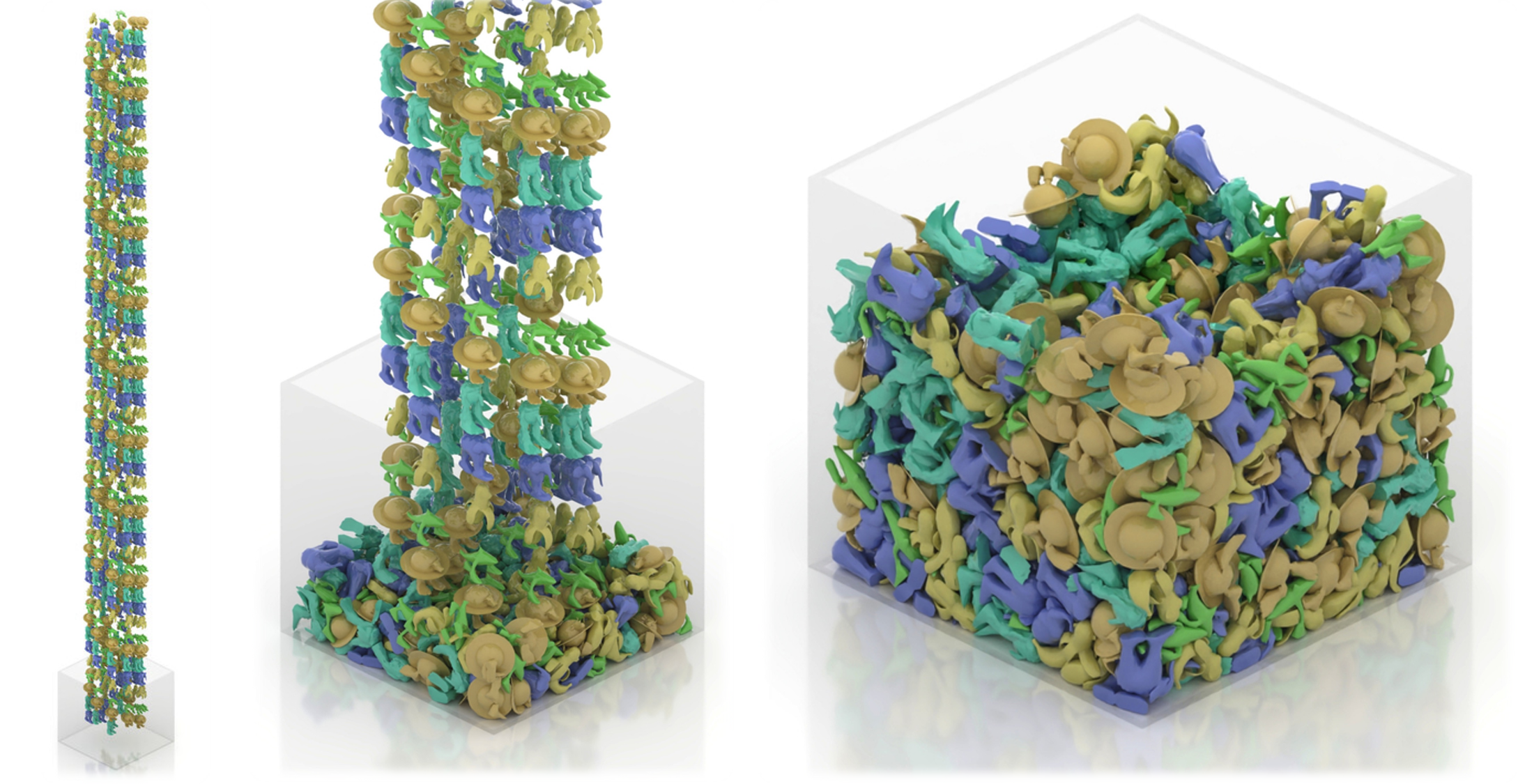}
  \caption{\textbf{Animal well.} A challenging test case featuring a large collection of objects (1.34M tetrahedra) and high velocity induced by gravity. The topmost objects are accelerated to $19.7\,\text{m/s}$ when colliding with the lower ones, traveling farther than their average size within a single time step.}
  \label{fig:animals}
\end{figure}

\paragraph{Augmented Lagrangian Formulation}
We then design a customized Augmented Lagrangian method by introducing explicit Lagrange multipliers to enforce constraints accurately without relying on excessively stiff penalty terms.

Specifically, we first introduce a set of non-negative slack variables $\mathbf s=\{s_i\in[0,\infty)\}_{i\in \mathcal I_\mathcal C}$ to transform the inequality constraints
$c_i(\hatx)\ge 0$
into equality constraints
\begin{equation} \label{eq:equality-constraint}
c_i(\hatx)-s_i=0,
\end{equation}
and then explicitly maintain a set of estimated Lagrange multipliers $\lambda_i$ and add a Lagrangian term and a penalty term to the IP:
\begin{equation} \label{eq:total-energy}
\mathcal L:=E(\hatx,\tildex)+\sum_{i\in \mathcal I_\mathcal C}\gamma_i\left(\frac{\mu}2(c_i(\hatx)-s_i)^2-\lambda_i(c_i(\hatx)-s_i)\right).
\end{equation}
Here, $\mu$ is the penalty stiffness and $\gamma_i$ are the decay factors to be discussed later. The standard Augmented Lagrangian method guarantees that if we alternate between minimizing $\mathcal L$ w.r.t. the primal variables $(\hatx,\mathbf s)$ and updating the multipliers via
\begin{equation} \label{eq:lambda_update}
\lambda_i\gets\lambda_i-\mu(c_i(\hatx)-s_i),
\end{equation}
then with a reasonably large fixed $\mu$, the multipliers $\lambda_i$ will eventually converge to the true Lagrange multipliers satisfying the KKT conditions \cite{nocedal2006numerical}. Subsequently, $\hatx$ will also converge to the solution of \autoref{eq:sub-problem} up to a small contact gap $\delta$ with $c_i(\hatx)-s_i$ tending to zero.

\paragraph{Alternating Primal Solve}
However, jointly optimizing $(\hatx,\mathbf s)$ remains a challenging inequality constrained optimization problem. In addition, without sufficiently accurate $\lambda_i$, solving this problem to high accuracy may be unnecessary. After exploration, we observed that a more practical strategy is sufficient: we further alternate between optimizing the slack variables via
\begin{equation}
s_i\gets \max(0,c_i(\hatx)-\lambda_i/\mu)
\end{equation}
and updating $\hatx$ by projected Newton iterations with line search to ensure decreasing total energy $\mathcal L$. We terminate this inner iteration when a full step is taken in the line search for $\hatx$. For any elasticity linear in $\hatx$, it always terminates in a single iteration. For nonlinear and non-convex elasticities, it effectively prevents $\lambda_i$ from blowing up when $\hatx$ becomes trapped at irregular landscapes, while still terminating in one iteration in most cases.

\begin{algorithm}
\caption{SolveSubproblem}\label{alg:subproblem}
\begin{algorithmic}[1]
\Require inertia target $\tilde{\mathbf x}$, intersection-free state $\mathbf x$, initial guess $\hat{\mathbf x}^0$, current active set $\mathcal C=\{(i,\lambda_i,\gamma_i)\}_{i\in\mathcal I_\mathcal C}$.
\Ensure approximate solution $\hat{\mathbf x}$ for \autoref{eq:sub-problem}.
\State $\hatx\gets\hatx^0$;
\ForAll{$(i,\lambda_i,\gamma_i)\in\mathcal C$}
\State Linearize the unsigned distance using $d_i(\x)$ and $\nabla d_i(\x)$;
\EndFor
\While{true}
\State $\mathbf G\gets\nabla_{\hatx} E(\hatx,\tildex)$;
\State $\mathbf H\gets$ \Call{SPDProject}{$\nabla^2_{\hatx}E(\hatx,\tildex)$};
\ForAll{$(i,\lambda_i,\gamma_i)\in\mathcal C$}
\State $s_i\gets\max(0,c_i(\hatx)-\lambda_i/\mu)$;
\State $\mathbf G\gets\mathbf G+\mu\gamma_i(c_i(\hatx)-\lambda_i/\mu-s_i)\nabla d_i(\x)$;
\State $\mathbf H\gets\mathbf H+\mu\gamma_i\nabla d_i(\x)\nabla d_i(\x)^T$;
\EndFor
\State $\mathbf p\gets-\mathbf H^{-1}\mathbf G$;\Comment{PCG Solve}
\State $r\gets$ \Call{LineSearch}{$\hatx,\mathbf p$};\Comment{$\mathcal L(\hatx+r\mathbf p)<\mathcal L(\hatx),0\le r\le 1$}
\State $\hatx\gets\hatx+r\mathbf p$;
\If{$r=1$}
\State break;
\EndIf
\EndWhile
\ForAll{$(i,\lambda_i,\gamma_i)\in\mathcal C$}
\State $s_i\gets\max(0,c_i(\hatx)-\lambda_i/\mu)$;
\If{$s_i=0$}
\State $\lambda_i\gets\lambda_i-\mu c_i(\hatx)$;\Comment{Equivalent to \autoref{eq:lambda_update}}
\State $\gamma_i\gets1$;
\Else
\State $\lambda_i\gets0$;\Comment{Equivalent to \autoref{eq:lambda_update}}
\State $\gamma_i\gets\Gamma\gamma_i$;
\EndIf
\EndFor
\State \Return $\hatx$;
\end{algorithmic}
\end{algorithm}

\paragraph{Constraint Decay}
\label{sec:decay}
Algorithm \ref{alg:subproblem} shows our solver for the subproblem (\autoref{eq:sub-problem}). Line 8 and 9 compute the gradient and Hessian of the augmented penalty terms, which can be easily derived from \autoref{eq:total-energy}. Notably, in our Newton update for the primal variables, the constraints with $s_i>0$ will not affect the gradient $\mathbf G$ since $c_i(\hatx)-\lambda_i/\mu-s_i=0$, but they will still generate non-zero terms in the Hessian $\mathbf H$. For these constraints in $\mathcal{C}$ that are actually inactive, the additional Hessian terms may slow down the convergence and performance of the subproblem solver. However, instantly removing these constraints from $\mathcal C$ will easily lead to oscillation and stability issues, as we will show in \S \ref{sec:ablation}. Our solution is to introduce the decay factors $\gamma_i$ to smoothly reduce the influence of the inactive constraints and restore them once they become active again. Specifically, $\gamma_i$ is multiplied by a decay factor $\Gamma$ per iteration when the constraint is inactive, scaling down the contribution of the augmented Lagrangian terms in \autoref{eq:total-energy}. Constraints that remain inactive for multiple iterations are eventually removed, as detailed in the next subsection. We use $\Gamma = 0.9$ in all experiments, which we find to be the most effective setting (see \S\ref{sec:ablation}).

\subsection{Active Set Update}
\label{sec:activeset}

Another challenge remaining in our framework is to effectively update the active set $\mathcal C^{[k]}$ to ensure large step size $\alpha^{[k]}$ while restricting the number of active constraints. When advancing $\x^{[k]}$ to $\hatx^{[k+1]}$ using CCD, if the primitive pairs previously blocking the CCD from $\x^{[k-1]}$ to $\hatx^{[k]}$ are not resolved, they will persist producing near-zero collision times in the new CCD queries, thereby resulting in tiny $\alpha^{[k+1]}$. Therefore, a natural way to update the active set is directly adding all primitive pairs generating intersections during the last CCD into $\mathcal C^{[k+1]}$.

\begin{algorithm}
\caption{UpdateActiveSet}\label{alg:activeset}
\begin{algorithmic}[1]
\Require intersection-free state $\mathbf x$, target state $\hat{\mathbf x}$, current active set $\mathcal C=\{(i,\lambda_i,\gamma_i)\}_{i\in\mathcal I_\mathcal C}$.
\Ensure updated active set $\mathcal C'$.
\State $\mathcal C'=\mathcal C$;
\State Collect all primitive pairs $\mathcal I'$ that generate intersections during transition from $\x$ to $\hatx$;\Comment{BVH \& CCD queries}
\State $\mathcal I'\gets\mathcal I'\setminus \mathcal I_\mathcal C$;
\State \textbf{for all} $i\in\mathcal I',T_i\gets$ CCD collision time of primitive pair $i$;
\ForAll{vertex $v$}
\State $T_v\gets\min_{i\in \mathcal I'\text{ and } v\in i}T_i$;
\EndFor
\ForAll{$i\in \mathcal I'$}
\If{$T_i \in \{ T_v \ | \ v\in i\}$}
\State $\lambda_i,\gamma_i\gets0,1$;
\State $\mathcal C'\gets\mathcal C'\cup\{(i,\lambda_i,\gamma_i)\}$;
\EndIf
\EndFor
\ForAll{$(i,\lambda_i,\gamma_i)\in\mathcal C'$}
\If{$\gamma_i<0.01$}
\State $\mathcal C'\gets\mathcal C'\setminus\{(i,\lambda_i,\gamma_i)\}$;
\EndIf
\EndFor
\State \Return $\mathcal C'$;
\end{algorithmic}
\end{algorithm}

This simple strategy proves ineffective in scenarios involving large deformations and extremely complex contacts. Large variations between adjacent timesteps caused by large $h$ or velocity often leads to severe penetration in the intermediate states $\hatx^{[1]},\hatx^{[2]},\cdots$ during the early iterations. Under severe penetration, one primitive may pass through multiple layers of surfaces during the CCD, thereby generating far more active constraints than the actual ones. These unnecessary constraints may significantly interfere with the optimization process, leading to conflicting constraints and increased computational cost.

We aim to address this issue by filtering the constraints generated by the simple update strategy. For each vertex $v$, we first find the primitive pair with the earliest collision time among the newly generated constraints containing vertex $v$. We then keep only those primitive pairs corresponding to the earliest collision pair of at least one vertex. This filtering strategy does not lead to stagnation caused by missing essential constraints, as any primitive pair consistently detected by CCD will eventually pass the filtering and enter the active set. As shown in \S \ref{sec:ablation}, the filtering significantly reduces the number of active constraints, while effectively contributing to advancing the intersection-free trajectory. Algorithm \autoref{alg:activeset} details the subroutine for our active set update. The newly generated constraints are assigned $\lambda_i=0$ and $\gamma_i=1$. As stated in \S \ref{sec:subproblem}, decay factors of inactive constraints will be progressively reduced, and these constraints will be removed once the decay factor falls below the threshold value $0.01$.

\begin{figure}
\includegraphics[width=\linewidth]{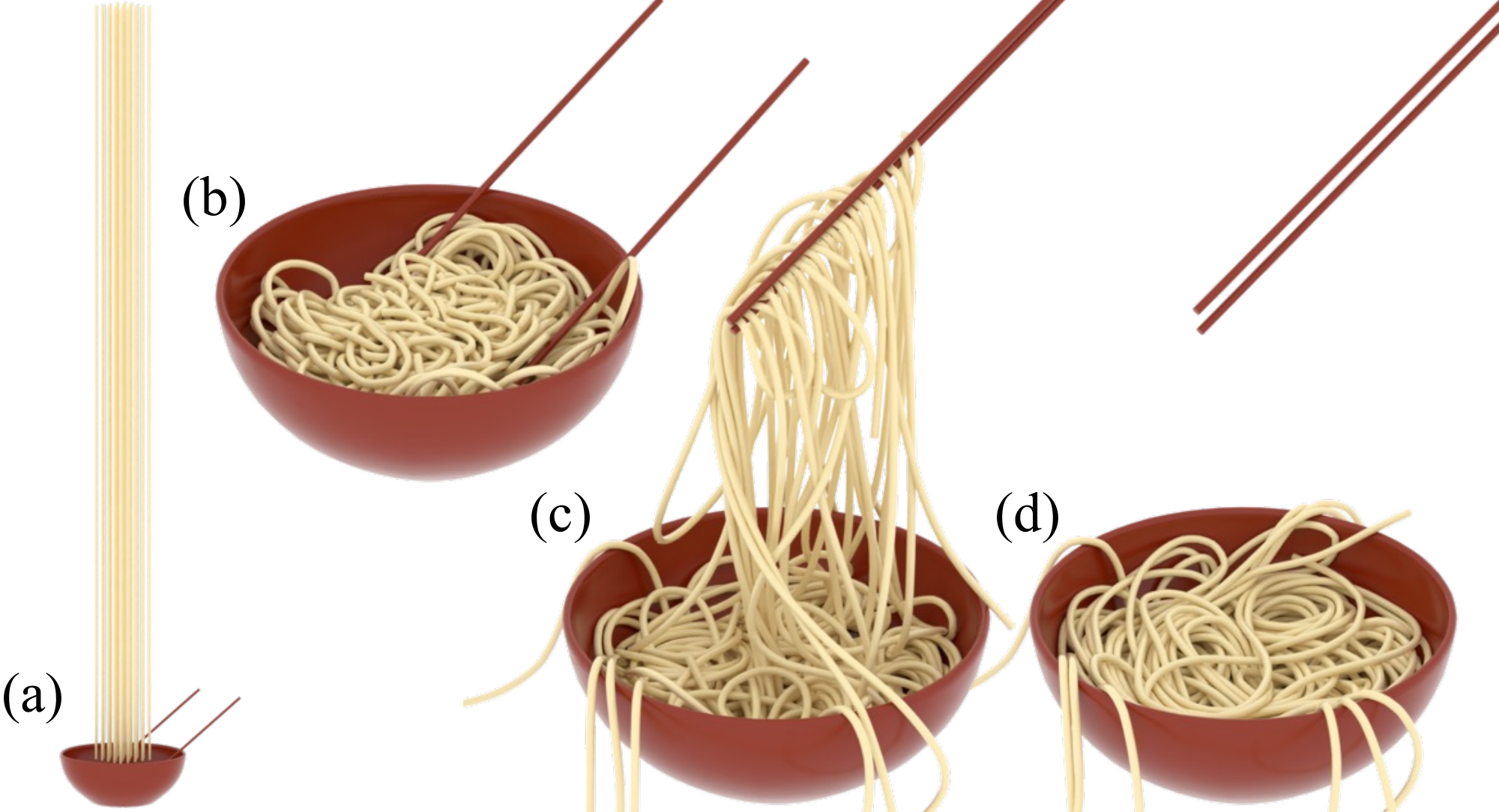}
  \caption{\textbf{Ramen.} An array of long noodles is dropped into a fixed bowl, stably picked up using chopsticks with static friction ($\mu_f=0.1$), and then released.}
  \label{fig:ramen}
\end{figure}

\subsection{Termination}
\label{sec:termination}

In \S\ref{sec:method-overview}, we described the TOI-based termination criterion (\autoref{eq:termination}) employed by our method. We now provide the high-level intuition that using a sufficiently small threshold $\epsilon$ and setting the minimum iteration count $K_\text{min} = 1$ offers a minimal yet effective condition for achieving high-fidelity simulation results. Similarly, we show that, under linear elasticity, this criterion leads to finite-step termination and first-order accurate time integration of our method.


\paragraph{Preserving Motion in Non-Contacting Regions}
Using a sufficiently small $\epsilon$ and setting $K_\text{min} = 1$ helps prevent the motion of non-contacting regions from being inadvertently constrained by time-of-impact (TOI) clamping in contact-rich areas. Consider a system with a contact-rich region and an isolated particle moving at constant velocity $\mathbf v$. Upon termination, the particle's net displacement over the timestep is $(1 - \beta_0^{[k]}) h \mathbf v$, which depends on the step sizes $\alpha^{[1]}, \dots, \alpha^{[k]}$ given by CCD in the contact region. By enforcing $\beta_0^{[k]} < \epsilon$ in the termination criterion, the particle retains at least $(1 - \epsilon)$ of its intended velocity, thereby avoiding artificially damped motion. This intuition is also provided in \cite{ando2024cubic}. Extending this reasoning, one can argue that non-contacting regions of elastic solids attain the same level of accuracy as semi-implicit Euler time integration.

\paragraph{Finite-Step Termination.}
To satisfy the TOI-based termination criterion in \autoref{eq:termination}, our method needs to avoid stagnation at $\alpha^{[k]} \rightarrow 0$. When $\alpha^{[k]}$ remains lower-bounded by a strictly positive value $\gamma>0$, each $\beta_i^{[k]}$ will become a product of values in $[0,1-\gamma)$ and converge to zero, ensuring eventual termination. The generation of such $\alpha^{[k]}$ can be achieved by the combination of our active set update and augmented Lagrangian framework. If $\alpha^{[k]}$ temporarily approaches zero, both $\x^{[k]}$ and the linearized constraints $\mathbf{c}(\hatx^{[k]})$ remain nearly unchanged during the outer iteration. In this case, Algorithm~\ref{alg:activeset} adds all primitive pairs immediately responsible for the tiny $\alpha^{[k]}$ into the active set $\mathcal{C}^{[k]}$ for subsequent iterations.
According to Theorem~17.6 of \citet{nocedal2006numerical}, when the elasticity model is linear and the penalty stiffness $\mu$ is sufficiently large, the errors of both the primal and dual variables relative to the true KKT solution of \autoref{eq:sub-problem} decay exponentially. Consequently, once the newly added contact pairs are included, the corresponding linearized constraints 
$
c_i(\hatx^{[k+1]}) = d_i(\x^{[k]}) + \nabla d_i(\x^{[k]})^T(\hatx^{[k+1]} - \x^{[k]}) - \delta
$
converge toward zero during the next subproblem solve. With $\delta > 0$, the resulting update direction $(\hatx^{[k+1]} - \x^{[k]})$ eventually aligns with the positive gradient directions of all $d_i(\x^{[k]})$, thereby producing a sufficiently large $\alpha^{[k+1]}$ in the next CCD.

\paragraph{First-order Accuracy.} We first examine the case where no contact exists in the scene, in which our solver is simply equivalent to performing $K_\text{min}$ Newton iterations. Let $\V^{[k]}=(\x^{[k]}-x^t)/h$, when we set $K_\text{min}=1$ (also known as semi-implicit Euler in some literature),

\begin{equation} \label{eq:semi-implicit}
\begin{aligned}
\V^{[k+1]}&=(\x^{[k]}-\frac{M(\x^{[k]}-\x^t-h\V^t)+h^2\nabla U(\x^{[k]})}{M+h^2\nabla^2U(\x^{[k]})}-\x^t)/h\\
&=\V^{[k]}-\frac{M(\V^{[k]}-\V^t)+h\nabla U(\x^{[k]})}{M+O(h^2)}\\
&=\V^t-hM^{-1}\nabla U(\x^t+h\V^{[k]})+O(h^2) \quad \text{by Neumann series}\\
&=\V^t-hM^{-1}\nabla U(\x^t)+O(h^2),
\end{aligned}
\end{equation}
where we denote $A/B=B^{-1}A$ for matrices $A$ and $B$, and assume Lipschitz continuity of the internal force $\nabla U$. Since, in the 2nd term here, $-M^{-1}\nabla U(\x^t)$ is the time derivative of $\V^t$, we know that the local truncation error of $\V^{[k+1]}$ (and similarly for $\x^{[k+1]}$) is $O(h^2)$ for any finite $k$.

When contact is present, we first consider time intervals during which the active constraint set remains unchanged. With sufficiently large $\mu$ and small $\delta$, our solver is equivalent to performing semi-implicit Euler (\autoref{eq:semi-implicit}) within the linear subspace satisfying all active constraints, and thus retains an $O(h^2)$ local truncation error. When the active constraint set changes, they happen momentarily, and assuming a smooth object trajectory with $h\rightarrow0$, there are only a finite number of such moments. Thus, even if our method needs several time steps to fully include all active constraints and remove inactive ones, the accumulated error is $O(1)O(h)$, since each such time step would introduce an $O(h)$ error in velocity due to the incorrect active set. Combined with the accumulated per time step error analyzed in \autoref{eq:semi-implicit}, the total accumulated error becomes $O(1)O(h)+O(h^{-1})O(h^2)=O(h)$, suggesting that our method is first-order accurate as a time integrator. We further validate this first-order accuracy experimentally in \S\ref{sec:unit-test} using both a 1D contact problem with an analytical solution (\autoref{fig:convergence}) and a 3D contact problem (\autoref{fig:balls_collide}).
\\

Our goal here is to provide a high-level intuition on how our method works, and we have not thoroughly considered residual errors in the momentum equation, nonlinear elasticity, etc., in our discussion.
In practice, we observed that Algorithm \ref{alg:subproblem} with line search is sufficient to robustly and effectively handle the nonlinear and non-invertible elasticities by directly solving the nonlinear programming problem.

In \S \ref{sec:ablation}, we demonstrate that the TOI-based termination criterion can also be applied to IPC, improving its performance without introducing damping artifacts. Nonetheless, our method remains significantly faster overall, owing to a higher average TOI and improved conditioning.

\section{SIMULATOR DESIGN AND OPTIMIZATION}

\subsection{GPU Optimization} \label{sec:gpu_opt}

All core components of our simulator, including collision detection, Hessian assembly, linear system solving, and state updates, are GPU-parallelized through CUDA kernels. Our implementation utilizes Thrust and cuBLAS for efficient GPU data management and linear algebra operations. We store the symmetric Hessian matrix in a sparse $3\times3$ Block Sparse Row (BSR) format, where only the diagonal and upper-triangular blocks are assembled. The linear systems are solved using the conjugate gradient method with a $3\times 3$ block-Jacobi preconditioner, where the symmetric BSR SpMV is accelerated via warp-level reduction as in \citet{huang2024stiffgipc}, which significantly reduces writing conflicts caused by atomic operations. The collision detection is accelerated using Linear BVH \citep{karras2012maximizing} in the broad phase and filtered with ACCD \citep{li2020codimensional} in the narrow phase.

\paragraph{Conflict-Free Accumulation of Analytic PSD Elasticity Hessians} In addition to collision detection and linear system solving, another time-consuming part of elastodynamic simulators on the GPU lies in the positive semidefinite (PSD) projection and assembly of the Hessian matrix, during which the local Hessian matrices of tetrahedral elements are projected onto the PSD region, and then assembled into the global Hessian in BSR format. Computing PSD projection of the local elasticity Hessian can easily become a bottleneck if iterative algorithms are used for numerical eigendecomposition.
A typical way to speedup this process is to analytically compute the eigenvalues $\omega_k$ and the corresponding eigenvectors $\text{vec}(\mathbf Q_k)$ of the $9\times9$ Hessian with respect to the deformation gradient $\mathbf F$. For isotropic elastic energies, these eigenvectors take the form
\begin{equation}
\mathbf Q_k=\mathbf U \mathbf D_k(\mathbf \Sigma)\mathbf V^T,\quad k=0,1,...,8,
\end{equation}
where $\mathbf F = \mathbf U\mathbf \Sigma\mathbf V^T$ is the SVD of $\mathbf F$ \citep{kim2020dynamic,smith2018stable}. The matrices $\mathbf D_k(\mathbf \Sigma)$ are diagonal for $0 \le k \le 2$ and antisymmetric with only two nonzero entries for $3 \le k \le 8$ (the $1/\sqrt2$ factor in some literature is absorbed into $\mathbf D_k$). A typical implementation (as in GIPC \citep{huang2024gipc}) of this analytical projection directly assembles the projected $9\times9$ Hessian
\begin{equation}
\left(\frac{\partial^2\Psi}{\partial\mathbf F^2}\right)^+=\sum_{k=0}^8\omega_k^+\text{vec}(\mathbf Q_k)\text{vec}(\mathbf Q_k)^T,\quad\omega_k^+=\max(\omega_k,0)
\end{equation}
and then transforms it back to the $12\times12$ form via two dense matrix multiplications. However, since the local assembly is parallelized at the tetrahedral element level, atomic operations are required when adding local matrices to the global Hessian, where we observe severe write conflicts that significantly limit performance. GIPC avoids write-conflict overhead by storing the Hessian in a matrix-free format that duplicates overlapping vertex blocks; however, this design increases the overhead of the subsequent SpMV operations in the PCG solve, which typically tends to be the more significant bottleneck in the simulation.

We propose a novel approach for improved GPU parallelization of the projection, which is parallelized at the vertex-block level to avoid write-conflict overhead via warp-level reduction, while also significantly reducing computation by exploiting the sparsity of $\mathbf D_k$. We first consider a $3\times3$ block in the previous $9\times 9$ Hessian:
\begin{equation}
\begin{aligned}
\left(\frac{\partial^2\Psi}{\partial\mathbf F^2}\right)_{i,j}^+&=\sum_{k=0}^8\omega_k^+\text{vec}(\mathbf Q_k)_{3i:3i+3}\text{vec}(\mathbf Q_k)_{3j:3j+3
}^T\\
&=\sum_{k=0}^8\omega_k^+(\mathbf U\mathbf D_k\mathbf V_i^T)(\mathbf U \mathbf D_k\mathbf V_j^T)^T\\
&=\mathbf U\left[\sum_{k=0}^8\omega_k^+\mathbf D_k(\mathbf V_i^T\mathbf V_j)\mathbf D_k^T\right]\mathbf U^T,
\end{aligned}
\end{equation}
where $0\le i,j\le 2$ are the block indices, and $\mathbf V_i,\mathbf V_j$ denote the corresponding rows of $\mathbf V$. When transforming back to $3\times 3$ block with respect to vertices $x_{i'}$ and $x_{j'}$, we sum up the $ij$-blocks by
\begin{equation}
\begin{aligned}
\left(\frac{\partial^2\Psi}{\partial x_{i'}\partial x_{j'}}\right)^+&=\sum_{0\le i,j\le 2}\mathbf A_{i',i}\mathbf A_{j',j}\left(\frac{\partial^2\Psi}{\partial\mathbf F^2}\right)_{i,j}^+\\
&=\mathbf U\left[\sum_{k=0}^8\omega_k^+\mathbf D_k(\mathbf A_{i'}\mathbf V)^T(\mathbf A_{j'}\mathbf V)\mathbf D_k^T\right]\mathbf U^T,
\end{aligned}
\end{equation}
where $\mathbf A_{i'}$ (resp. $\mathbf A_{j'}$) is the constant row vector mapping $dx_{i'}$ to $d\mathbf F$ (in an outer product $d\mathbf F = dx_{i'}\mathbf A_{i'}$), which can be precomputed at the beginning of the simulation. Before each Newton solve, we first compute the SVD and analytical eigendecomposition in parallel at the element level, storing $\mathbf U,\mathbf V,\omega_k$ for each element. We then parallelize at the vertex-block level; within each thread, we process an $i'j'$-block by first computing the outer product $(\mathbf A_{i'}\mathbf V)^T(\mathbf A_{j'}\mathbf V)$, then expanding the multiplications with $\mathbf D_k$ according to their sparse structures, and finally multiplying the result by $\mathbf U$ and $\mathbf U^T$. The threads are launched in the order of $i'$ and $j'$, so that all Hessian blocks corresponding to the same pair $(i', j')$ are aggregated via warp-level reduction, thereby alleviating atomic write conflicts. Compared to the direct analytical PSD projection, we achieve 2.34$\times$ fewer multiplications and 3.17$\times$ faster elasticity Hessian assembly. On our large-scale benchmark (\autoref{fig:animals}), the elasticity Hessian assembly still accounts for $11.6\%$ of the total runtime after optimization, highlighting the importance of this improvement.

\subsection{Conditioning-Aware Adjustment of $\mu$} \label{sec:adaptive_mu}

The theoretical termination guarantee of our method depends on sufficiently (though not infinitely) large $\mu$ and small $\delta$ in our augmented Lagrangian solver. An insufficiently small $\delta$ may cause conflicts among constraints, but fortunately, it can be set according to mesh resolution just like the distance threshold $\hat{d}$ in IPC. However, $\mu$ is not so intuitive, and setting it too large can easily make the system ill-conditioned. Thus, at the beginning of each time step, we estimate an initial value of $\mu$, setting it as large as possible while keeping the conditioning of $\nabla^2\mathcal{L}$ on the same level as that of $\nabla^2E$. In practice, we use the following approximation:
\begin{equation}
\mu_\text{init}=C\max_i\left(\nabla^2E\right)_{ii},
\end{equation}
which works well with a constant value of $C = 0.1$ used through all test cases.

\begin{figure}[h]
  \includegraphics[width=\linewidth]{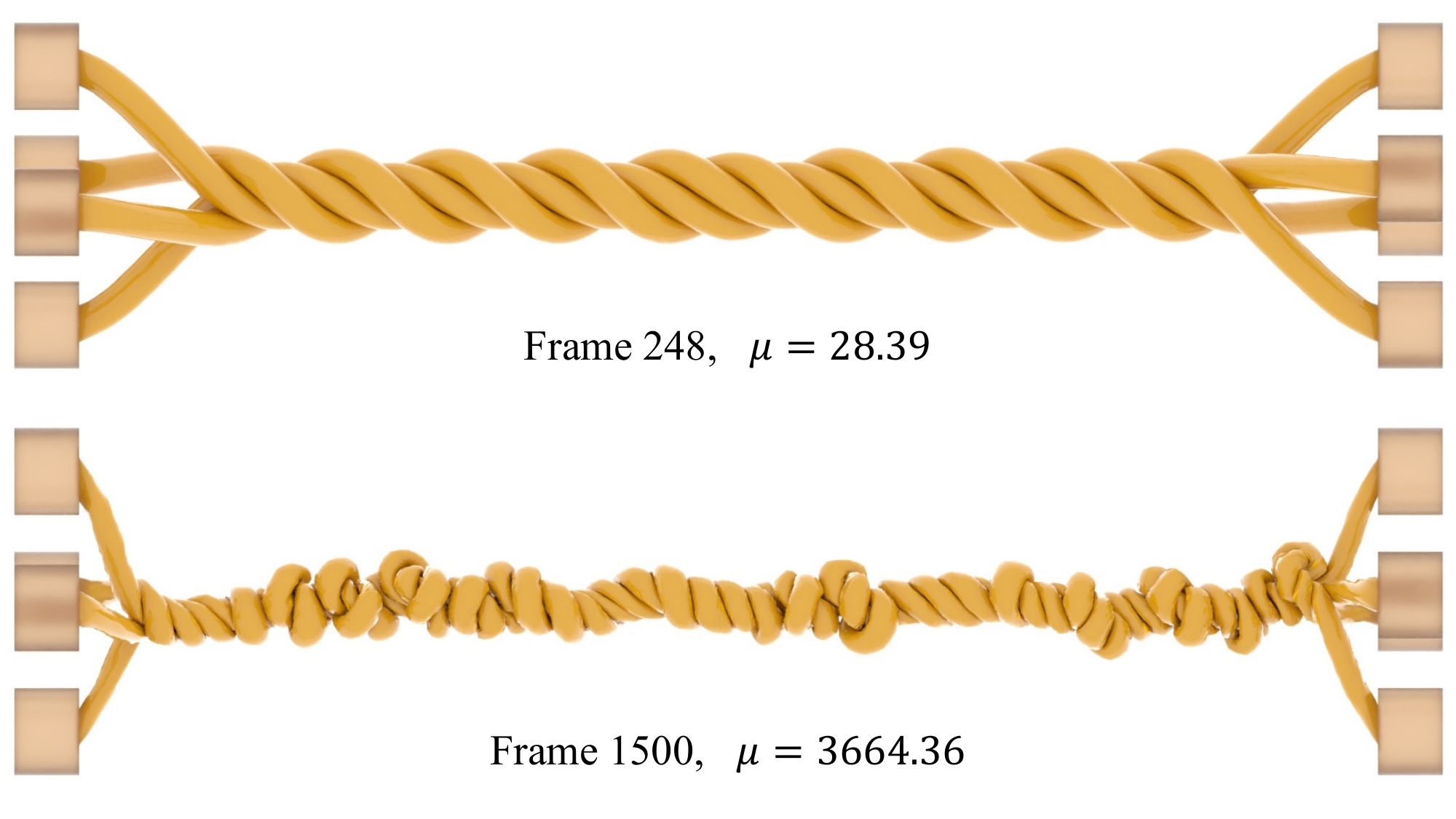}
  \caption{\textbf{Twisting rods.} A bundle of elastic rods with boundaries rotated by $5400^\circ$ in opposite directions at both ends. The contact stiffness $\mu$ is adaptively adjusted to match the conditioning of the Neo-Hookean elasticity.}
  \label{fig:rods}
\end{figure}

Theoretically, $\mu$ and $\delta$ should be adaptively adjusted when the iterations stall at zero TOI. We employ an adaptive mechanism that updates $\mu \gets 2\mu$ and $\delta \gets \delta / 2$ when $\alpha^{[k]} < 10^{-4}$ for 50 consecutive iterations; however, this condition was never triggered in any of our experiments, as the estimated $\mu_\text{init}$ already works effectively in practice. For example, in our twisting rods test (\autoref{fig:rods}), $\mu_\text{init}$ increases from \textasciitilde$10^1$ to \textasciitilde$10^4$ as the deformation becomes increasingly severe.


\subsection{Penalty-Free Moving Boundary Conditions}
\label{sec:DBC}

Another advantage of our method over barrier-based approaches is its ability to efficiently handle moving boundary conditions. In the IPC framework \citep{li2020incremental,huang2024gipc}, since the updated target positions of moving boundaries may penetrate other geometries, an auxiliary spring energy optionally augmented by a Lagrangian term is introduced to pull all moving boundary vertices toward their new positions, after which the moving boundary DoFs are eliminated from the system solve. This process requires additional iterations with poor conditioning due to the large spring stiffness. Our method avoids these extra iterations, as the intermediate states $\hatx^{[k]}$ allow penetration between the boundaries and interior geometries. In the initial state $\hatx^{[0]}$, we move all moving boundary vertices to their target positions and eliminate the moving boundary DoFs by zeroing out all corresponding entries in the system gradient and Hessian, except for the diagonals, in subsequent linear solves. Since all $\hatx^{[k]}$ satisfy the new boundary conditions, the boundaries in the final $\x^{[k]}$ is a linear blend between the target positions and the old boundaries in $\x^t$, with a blending weight on the target positions greater than $1-\epsilon$. The upper bound of the moving boundary position error is then
\begin{equation}
e_\text{boundary}\le \frac{\epsilon hv_\text{boundary}}{1-\epsilon},
\end{equation}
where $v_\text{boundary}$ denotes the maximum velocity of the moving boundary vertices. This provides a good trade-off between $e_\text{boundary}$ and efficiency.


\section{EVALUATION}

We conducted all evaluations on a desktop PC equipped with an Intel Core i9-13900K CPU (24 cores), 64 GB of RAM, and an NVIDIA GeForce RTX 4090 GPU. In all experiments, including the compared methods, we use the same CG relative error tolerance of $10^{-4}$ and a termination threshold of $\epsilon=10^{-3}$ for all methods employing TOI-based termination. We set $K_\text{min}=6$ for all cases involving friction and cloth and use $K_\text{min}=2$ otherwise (except Twisting cloth, see below). We present the unit tests in \S\ref{sec:unit-test}, stress tests in \S\ref{sec:stress-test}, ablation studies in \S\ref{sec:ablation}, and comparisons with state-of-the-art penetration-free simulators in \S\ref{sec:comparison}.

\subsection{Unit Tests} \label{sec:unit-test}

\begin{figure}
  \begin{subfigure}{0.49\linewidth}
        \includegraphics[width=\linewidth]{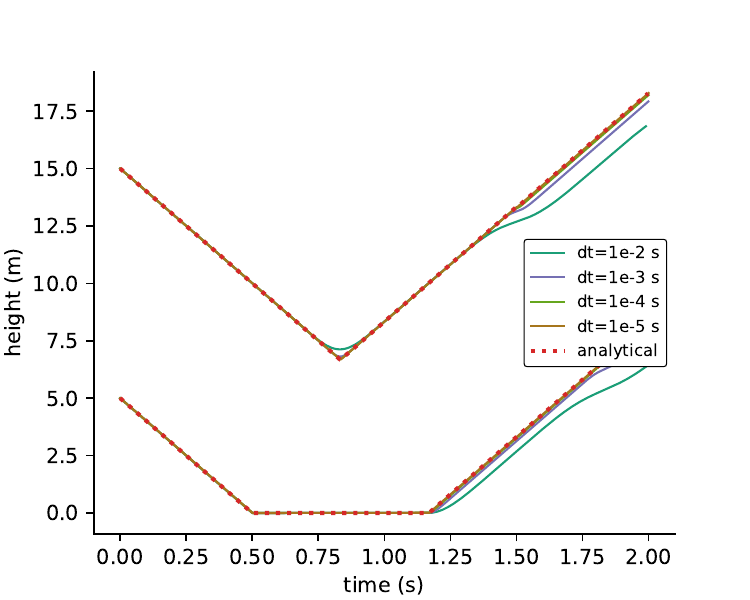}
        \caption{Problem 1 - w/o gravity}
        \label{fig:convergence_a}
    \end{subfigure}
    \begin{subfigure}{0.49\linewidth}
        \includegraphics[width=\linewidth]{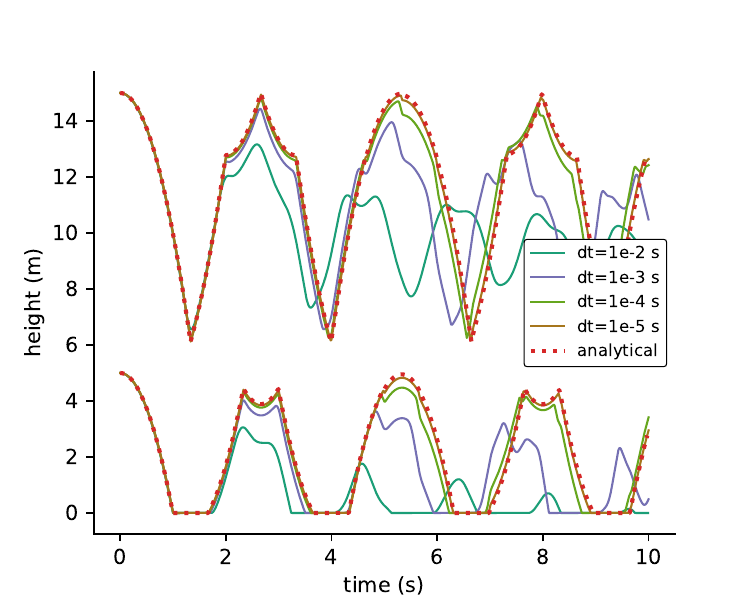}
        \caption{Problem 2 - w/ gravity}
        \label{fig:convergence_b}
    \end{subfigure}
    \begin{subfigure}{0.49\linewidth}
        \includegraphics[width=\linewidth]{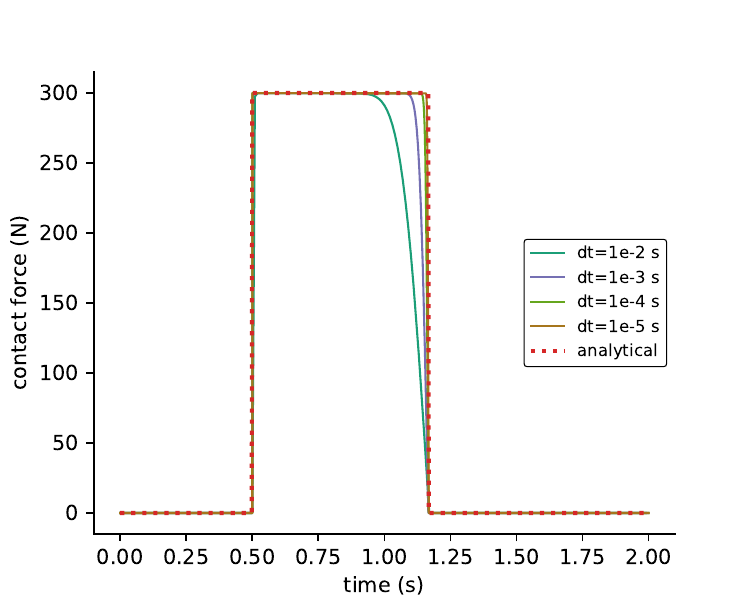}
        \caption{Contact Force}
        \label{fig:convergence_c}
    \end{subfigure}
    \begin{subfigure}{0.49\linewidth}
        \includegraphics[width=\linewidth]{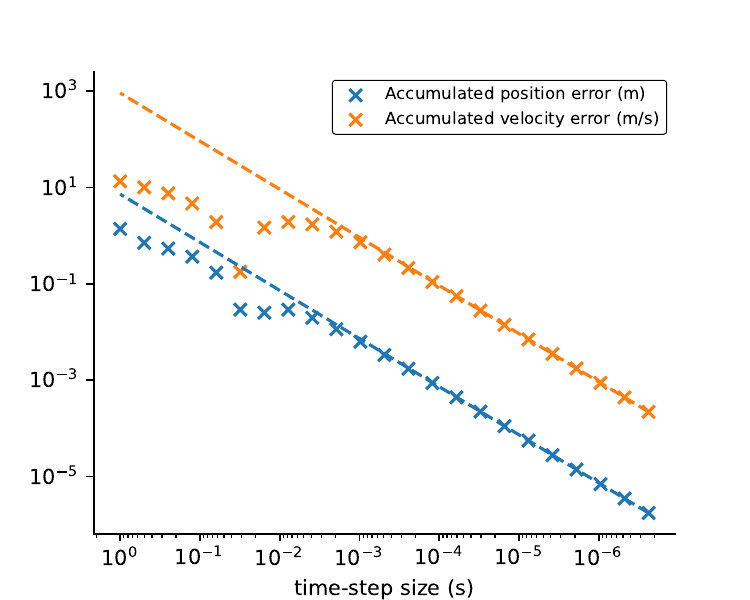}
        \caption{First-order Convergence}
        \label{fig:convergence_d}
    \end{subfigure}
  \caption{\textbf{First-order convergence: 1-dimensional contact.} As the time-step size is refined, our simulation results (heights of the elastic bar's endpoints) converge to the analytic solutions in both problems as shown in {(a)} and {(b)}. We further show for problem 1: {(c)} the convergence of contact force and {(d)} first-order convergence of accumulated position and velocity errors (the dashed reference lines are with slope 1).}
  \label{fig:convergence}
\end{figure}

\begin{figure}
  \includegraphics[width=\linewidth]{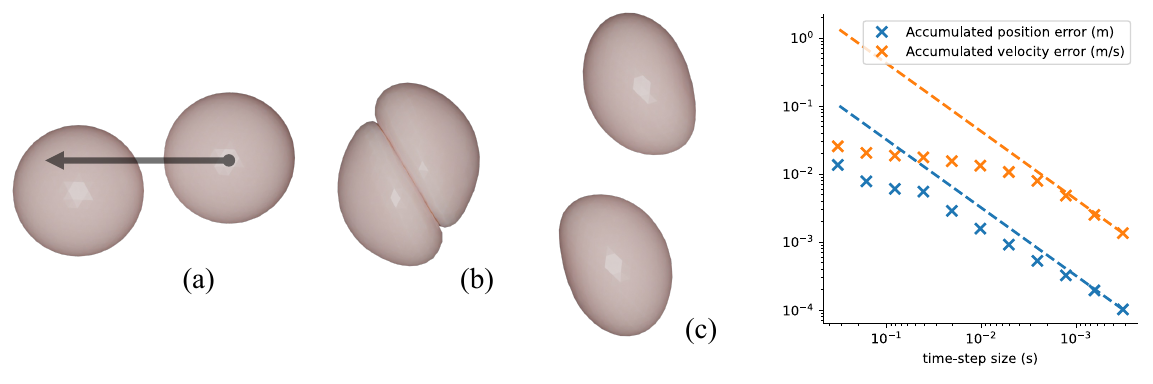}
  \caption{\textbf{First-order convergence: 3-dimensional contact.} Another unit test demonstrating first-order convergence of our method in the collision of two soft spheres. We plot the $L^2$ error norm against time step size, using $\Delta t=10^{-4}$s as the reference solution.}
  \label{fig:balls_collide}
\end{figure}

\paragraph{Convergence under time-step refinement} We first demonstrate the first-order time integration accuracy of our algorithm on two one-dimensional contact problems with known analytic solutions. Following \citet{doyen2011time, li2023convergent}, we test the collision between a linearly elastic bar and a rigid ground under two scenarios: (1) with an initial velocity but no gravity, and (2) with gravity but no initial velocity. We first calculate the exact solution for a fixed spatial discretization, and then progressively refine the time step sizes. As shown in \autoref{fig:convergence_a} and \ref{fig:convergence_b}, as the temporal discretization is refined, the simulated trajectories eventually converge to the analytic solutions. \autoref{fig:convergence_c} further shows the convergence of contact force between the objects in problem 1. The order of convergence is demonstrated in \autoref{fig:convergence_d}, where the accumulated position and velocity errors in problem 1 are plotted against time step sizes on a log-scale, with the data points following the reference lines with slope $1$. In \autoref{fig:balls_collide}, we demonstrate the first-order convergence of our method on a more complex three-dimensional contact problem involving the collision and detachment of two spheres. As no analytical solution is available, we use the simulation result with $\Delta t=10^{-4}$s as the reference and compute the averaged $L^2$ norm of the accumulated position and velocity errors.

\begin{figure}
  \includegraphics[width=\linewidth]{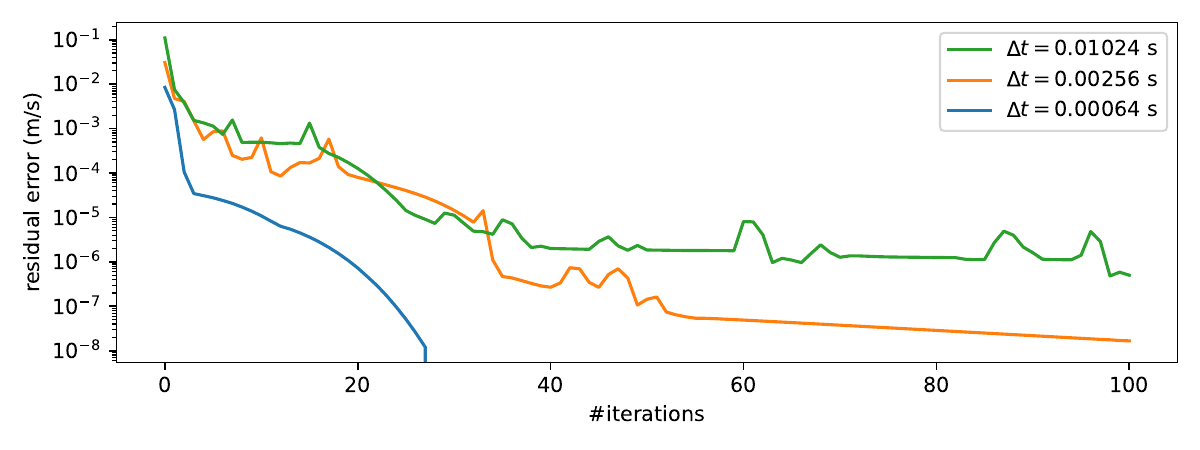}
  \caption{\textbf{Convergence of gradient-norm residual.} Convergence of our method on the IPC-style gradient norm-residual within a time step, starting from the system state shown in \ref{fig:balls_collide}(b).}
  \label{fig:balls_residual}
\end{figure}

\paragraph{Convergence of gradient-norm residual.} As in IPC, a residual related to the gradient $\nabla_\hatx \mathcal L$ can be defined to measure convergence toward a minimizer of $\mathcal L$. Although our pipeline employs a TOI-based termination criterion to improve efficiency while preserving simulation fidelity, we also demonstrate convergence under an IPC-style residual-based termination criterion in \autoref{fig:balls_residual}. Following \citet{li2020incremental}, we compute the Hessian-scaled gradient norm $\frac1{\Delta t}\|(\nabla^2\mathcal L)^{-1}\nabla\mathcal L\|_\infty$ and plot the convergence behavior under different $\Delta t$, starting from the system state shown in \autoref{fig:balls_collide}(b). The plot shows that our method can converge to high accuracy, driving the residual below $10^{-6}$, where IPC's default tolerance is $10^{-2}$. The convergence here exhibits a long tail, consistent with the 1st-order dual update steps in the augmented Lagrangian method. As the time step size decreases, the starting point ($\mathbf{x}^t$) gets closer to the solution and the active set evolves more smoothly, leading to faster convergence.

\begin{figure}
  \includegraphics[width=\linewidth]{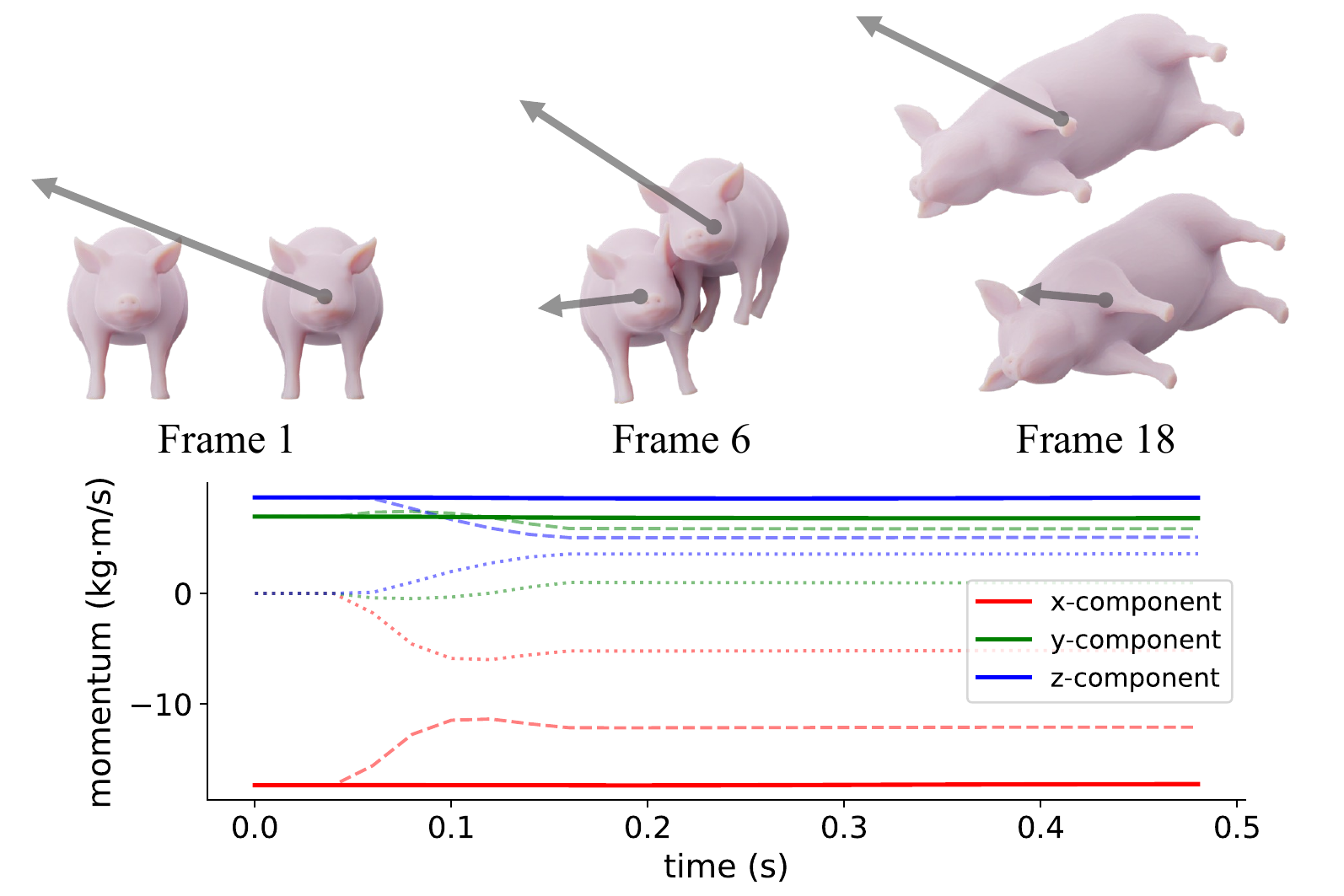}
  \caption{\textbf{Momentum conservation.} The plot shows the evolution of total and individual momenta, where solid lines represent the total momentum and dotted lines denote each body.}
  \label{fig:momentum}
\end{figure}

\paragraph{Momentum conservation} We validate the conservation of system's total linear momentum during frictional contact between two soft bodies, as shown in \autoref{fig:momentum}. We apply an initial velocity on one of the body, making it collide with the other body and transfer its momentum through contact and friction forces. After the two bodies separate, the system reaches a state of constant motion while preserving the same total momentum as in the initial configuration.

\begin{figure}
  \includegraphics[width=\linewidth]{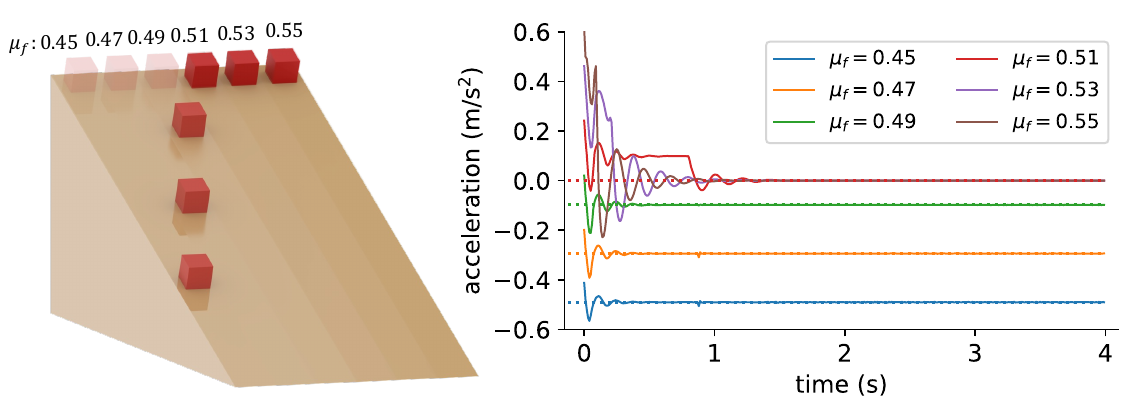}
  \caption{\textbf{Cubes on a slope.} Cubes placed on a $\arctan 0.5\approx26.6^\circ$ inclined surface with friction coefficients ranging from $0.45$ to $0.55$, demonstrating accurate frictional behavior near the sliding threshold.}
  \label{fig:slope}
\end{figure}

\paragraph{Sliding friction} We evaluate the accuracy of our friction model by placing initially stationary cubes on a fixed surface inclined at $\arctan 0.5\approx26.6^\circ$ (\autoref{fig:slope}). The friction coefficients between the cubes and the surface range from $0.45$ to $0.55$, covering the threshold value of $0.5$ that allows sliding. Our method accurately simulates the frictional behavior with friction coefficients near the threshold value, as shown in the acceleration curves versus analytic solutions.

\begin{figure}
  \includegraphics[width=0.75\linewidth]{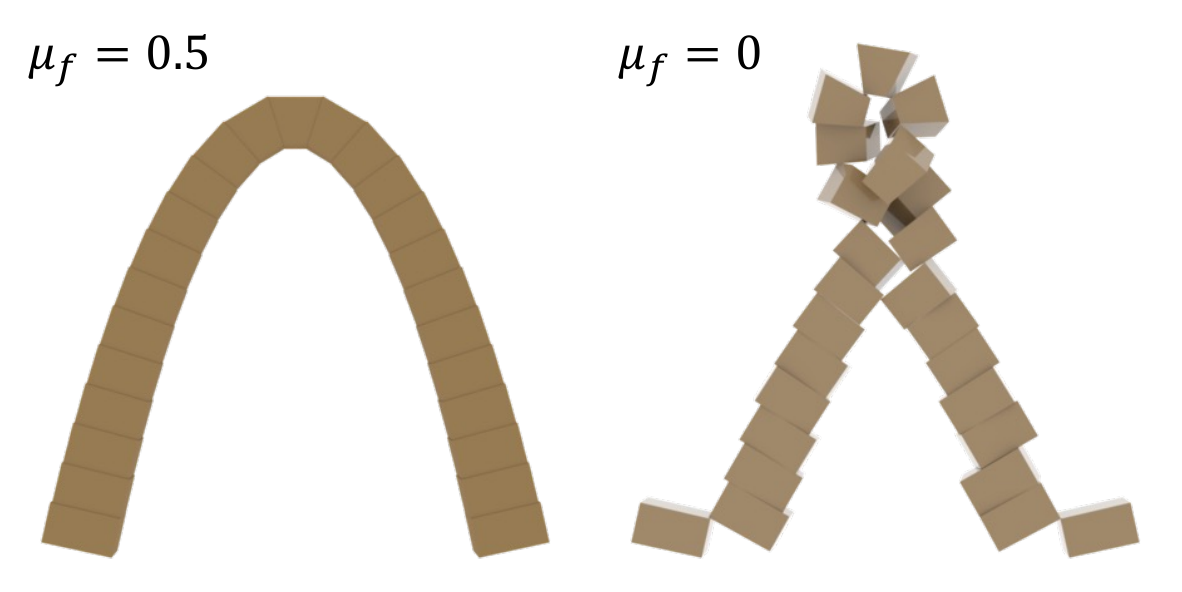}
  \caption{\textbf{Masonry arch.} Following \citet{li2020incremental}, we demonstrate robust static friction with coefficient $\mu_f = 0.5$ that supports the arch structure.}
  \label{fig:arch}
\end{figure}

\paragraph{Static friction} Following \citet{li2020incremental}, we include the masonry arch benchmark for static friction. We model the arch as a set of nearly rigid blocks with a Young’s modulus of $E = 10$ MPa, with the two bottom blocks fixed as boundary supports (\autoref{fig:arch}). We observe that the blocks form a stable arch structure under a friction coefficient of $\mu_f = 0.5$, while collapsing when $\mu_f = 0$ due to sliding between the blocks.

\begin{figure}
  \includegraphics[width=\linewidth]{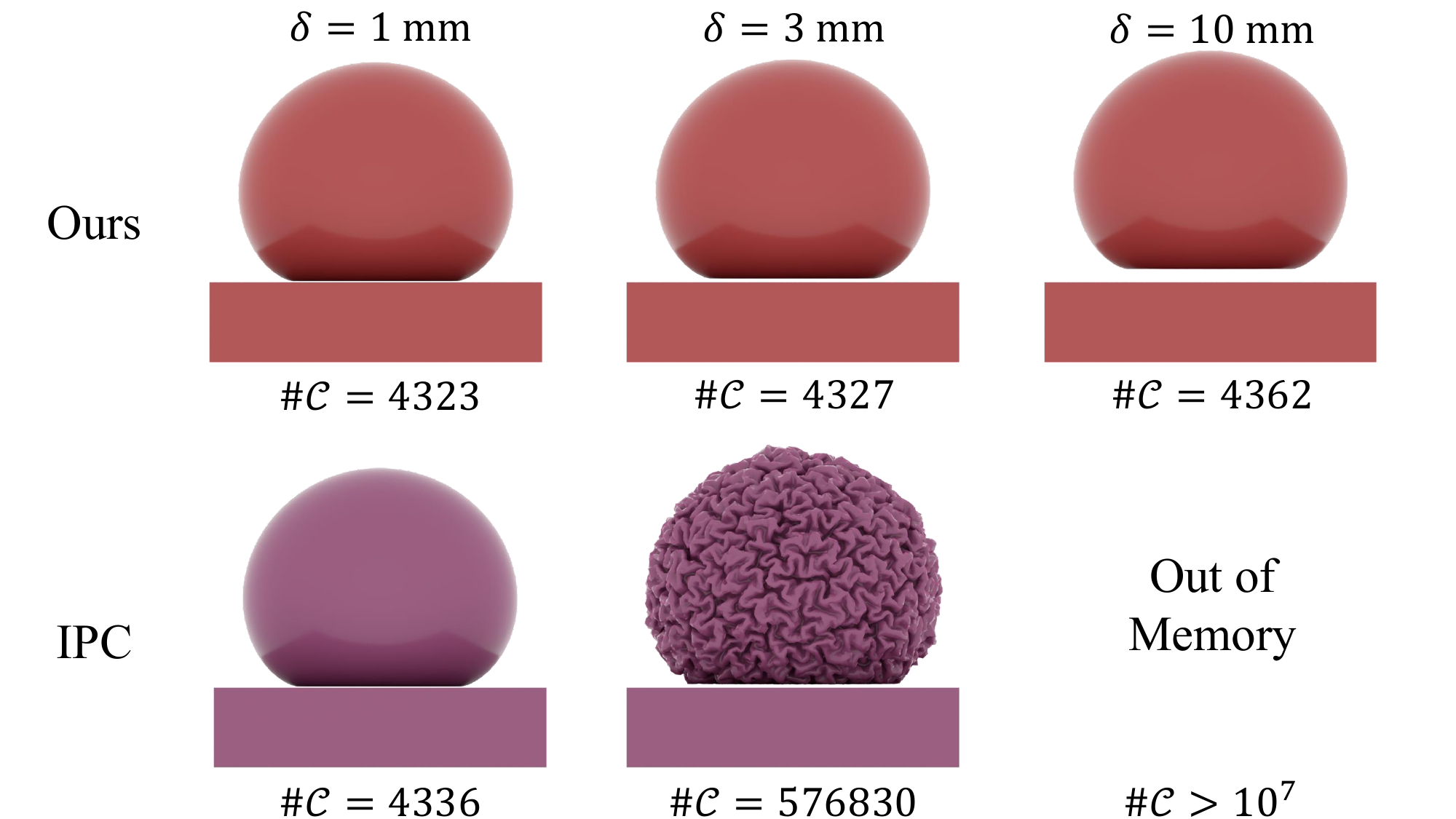}
  \caption{\textbf{High-resolution mesh under large contact radius.} Dropping a high-resolution sphere with 50k vertices. The number of contact pairs stays consistent in our method, while IPC generates a large number of false contact pairs and lead to severe artifacts (bottom middle) or runs out of memory (bottom right).}
  \label{fig:fine}
\end{figure}

\paragraph{High-resolution mesh under large $\delta$} A well-known issue \cite{huang2025geometric, chen2025offset} of IPC is that when the mesh resolution is sufficiently high that the surface triangle sizes are smaller than $\hat d$ (corresponding to $\delta$ in our method), the barrier penalty generates contact forces between geodesically close triangle pairs, which can lead to severe artifacts and a large number of false contact pairs (see the second row of \autoref{fig:fine}). In comparison, our method only adds colliding primitive pairs detected during CCD into the constraint set, and thus avoids generating contact pairs that are in fact not colliding. As shown in \autoref{fig:fine}, with increasing $\delta$, our method maintains a stable number of active constraints and does not suffer from the artifacts observed in IPC.

\subsection{Stress Tests} \label{sec:stress-test}

\begin{figure}
  \includegraphics[width=\linewidth]{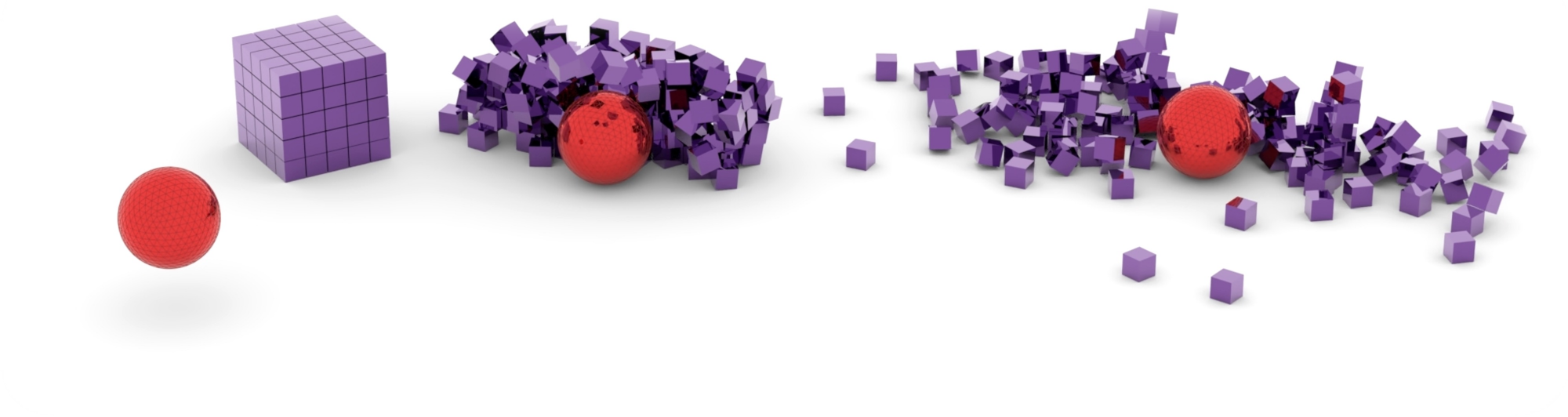}
  \vspace{-0.8cm}
  \caption{\textbf{Near-rigid bodies.} A near-rigid sphere with $E=1$ GPa collides with a pile of equally stiff cubes, generating intense collisions and rolling behavior induced by ground friction.}
  \label{fig:rigid}
\end{figure}

\paragraph{Near-rigid bodies} We evaluate the ability of our algorithm to simulate contact of near-rigid objects with high stiffness. As illustrated in \autoref{fig:rigid}, we simulate a moving sphere crashing into a pile of cubes, with all objects having a large Young's modulus of $E=1$ GPa. Our simulation remains stable and efficient during the high-speed impact and the complex contacts between the objects, while also nicely capturing the sphere's rolling behavior induced by ground friction (see our supplemental videos).

\begin{figure}
  \includegraphics[width=\linewidth]{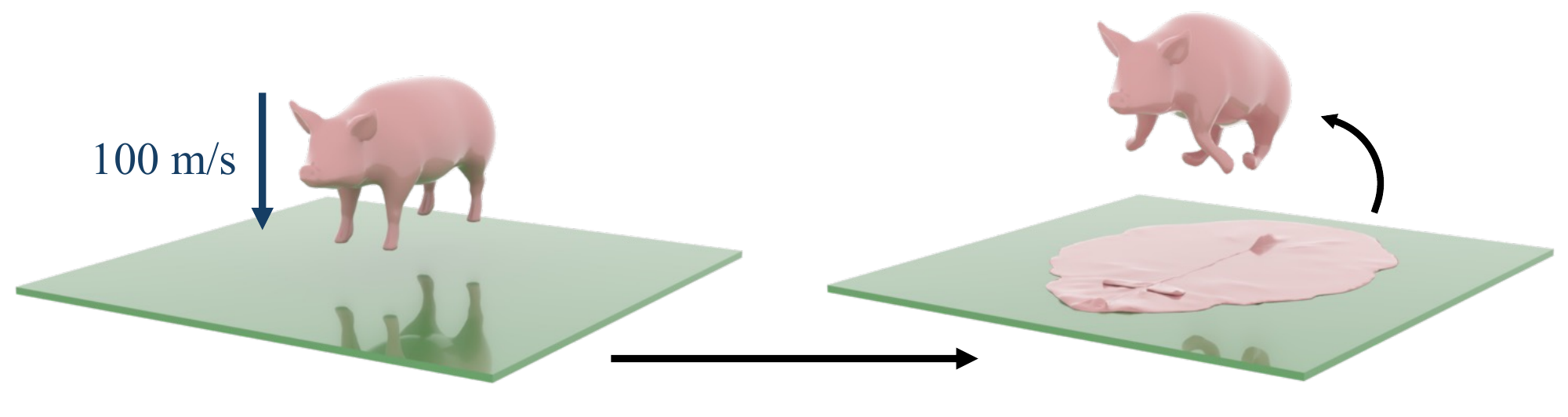}
  \caption{\textbf{Pig falling.} A pig falls onto a fixed thin plate at an extremely high speed (100 m/s), compressed into a thin layer, and bounces back to recover its rest shape. Our results remain penetration-free throughout the process.}
  \label{fig:fall}
\end{figure}

\paragraph{Extremely high-speed contact} We evaluate the robustness of our method using a challenging test case in which a soft body collides with a fixed thin plate at an extremely high speed of 100 m/s (\autoref{fig:fall}), a scenario that most methods lacking penetration-free guarantees will suffer from tunneling artifacts. During contact, the huge momentum results in an extreme compression of the body into a thin layer, after which it recovers its shape and bounces back. The whole process remains stable and penetration-free.


\paragraph{Squishy balls under extreme compression} We present a challenging test case to evaluate our method's robustness under complex contacts and extreme stress. As shown in \autoref{fig:teaser}, we simulate five elastic squishy balls compressed by a hydraulic press machine, modeled as a moving Dirichlet boundary. We first progressively shrink the internal space of the container, reaching a minimum height of $5$ mm, which results in an extreme compression of the squishy balls, generating $1.45M$ contact pairs at peak. We then quickly release the top plate, allowing the squishy balls to recover their shapes and rebound as a result of the stored elastic potential energy.

\paragraph{Animal well} To evaluate the scalability of our method, we simulate a scene with a large collection of elastic animal toys falling into a fixed square well, as illustrated in \autoref{fig:animals}. The system contains 1.29M DoFs and 1.34M tetrahedral elements, which generates up to 181.4k contact pairs at peak. Under gravitational acceleration, the objects at the top reach a maximum speed of $19.7\text{ m/s}$ when colliding with the lower bodies, such that the distance traveled in a single time step exceeds the body size. This again illustrates the robustness of our method under high-speed impacts and complex contact in high-resolution scenarios.

\paragraph{Compressing chains} We introduce another challenging example to demonstrate our method's robustness under extreme compression. As shown in \autoref{fig:chains}, three nested elastic chain rings with a Young's modulus of $E=100$ kPa are compressed within a shrinking box-shaped boundary. The volume occupied by the rings decreases from $0.04\,\text{m}^3$ to $215.9\,\text{cm}^3$ during compression, resulting in an average density increase of $185.2\times$. The boundary is then released, making the chains rapidly expand outward driven by the high elastic potential energy stored during compression. Our method robustly simulates the compression and expansion of the chains, preserving their exact topology even under extreme compression.

\paragraph{Ramen} As shown in \autoref{fig:ramen}, the Ramen test demonstrates complex frictional contacts, in which 25 ramen noodles ($12$ meters long, Young's modulus $E=32$ kPa) are dropped into a fixed bowl and subsequently picked up with a pair of chopsticks. With a friction coefficient of $\mu_f = 0.1$ between the chopsticks and the noodles, we are able to simulate stable static friction that prevents the noodles from sliding down. The gripped noodles naturally slip off when the gap between the two chopsticks increases. The simulation is efficiently time stepped at 0.33s per time step ($h=0.02s$).

\paragraph{Twisting rods} We follow \citet{li2020incremental} to introduce the example of a bundle of twisting rods (\autoref{fig:rods}), each with a stiff Young's modulus of $E=1$ MPa. With both ends rotating at $180^\circ$/s ($2.5\times$ faster than the original setting) in opposite directions for $30$ s, we generate large deformations on the four thin elastic rods and observe strong buckling effects under high stress.

\begin{figure}
  \includegraphics[width=\linewidth]{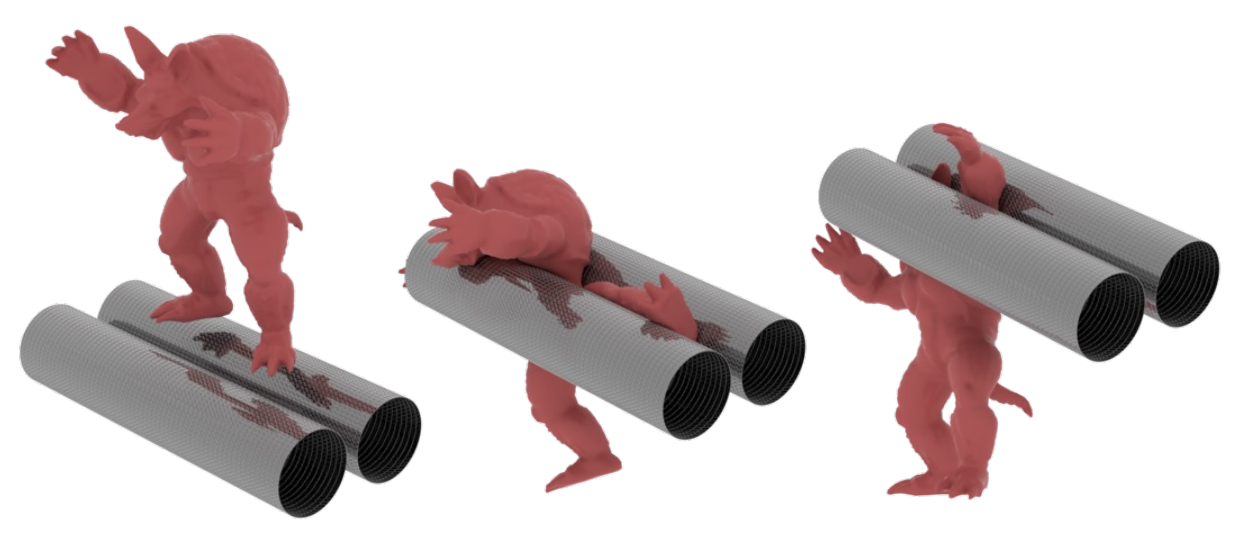}
  \caption{\textbf{Friction roller.} A stiff Armadillo model ($E=10$ MPa) is dropped onto two rotating rollers with friction coefficient $\mu_f=0.1$, demonstrating robust frictional contact under large deformation.}
  \label{fig:roller}
\end{figure}

\paragraph{Friction roller} Another test case (\autoref{fig:roller}) from \citet{li2020incremental} demonstrates the robustness of friction under large deformation. A stiff Armadillo model with Young's modulus of $E=10$ MPa is dropped onto a pair of fixed rollers rotating at a constant speed, with a friction coefficient of $\mu_f = 0.1$ applied between them. We robustly handle the static friction between the objects, which drives the Armadillo downward through the narrow gap against the large elastic resistance. The simulation proceeds efficiently with $h=0.02s$, achieving an average runtime of 0.15s per step, even under large deformations and frictional contact.  

\begin{figure}
  \includegraphics[width=\linewidth]{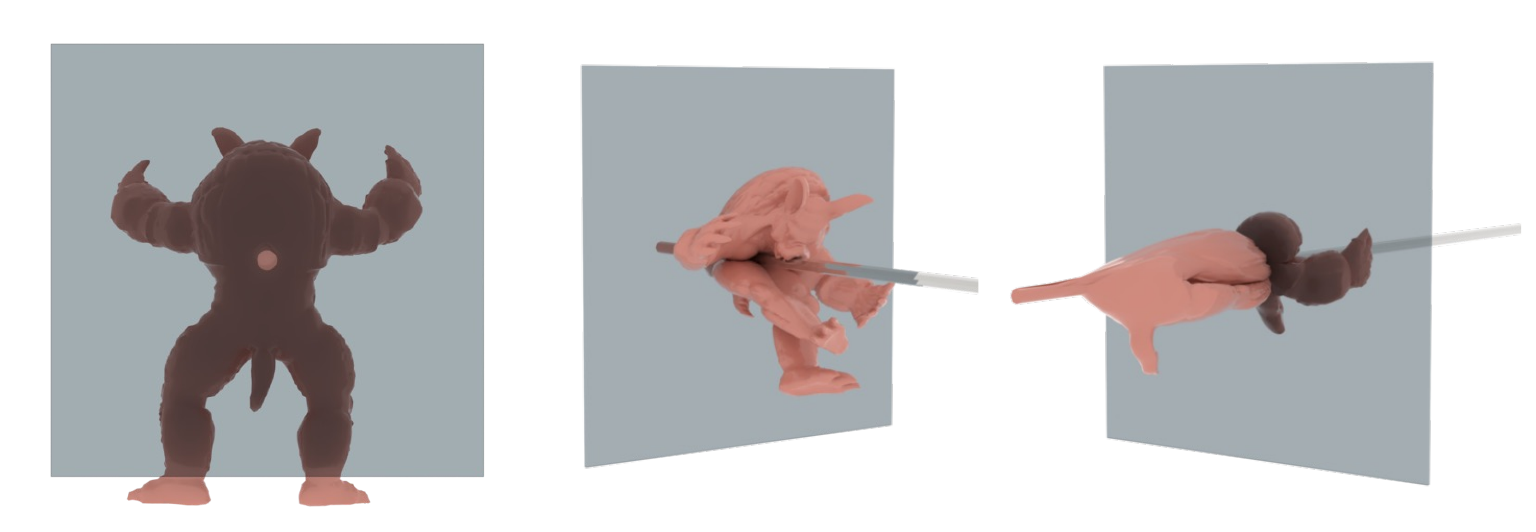}
  \caption{\textbf{Armadillo through a tiny hole.} A stiff Armadillo model ($E=10$ MPa, 10 m tall) is pushed through a narrow circular hole by a moving rod, demonstrating robust handling of extreme deformation and high stress.}
  \label{fig:hole}
\end{figure}

\begin{figure}
  \includegraphics[width=\linewidth]{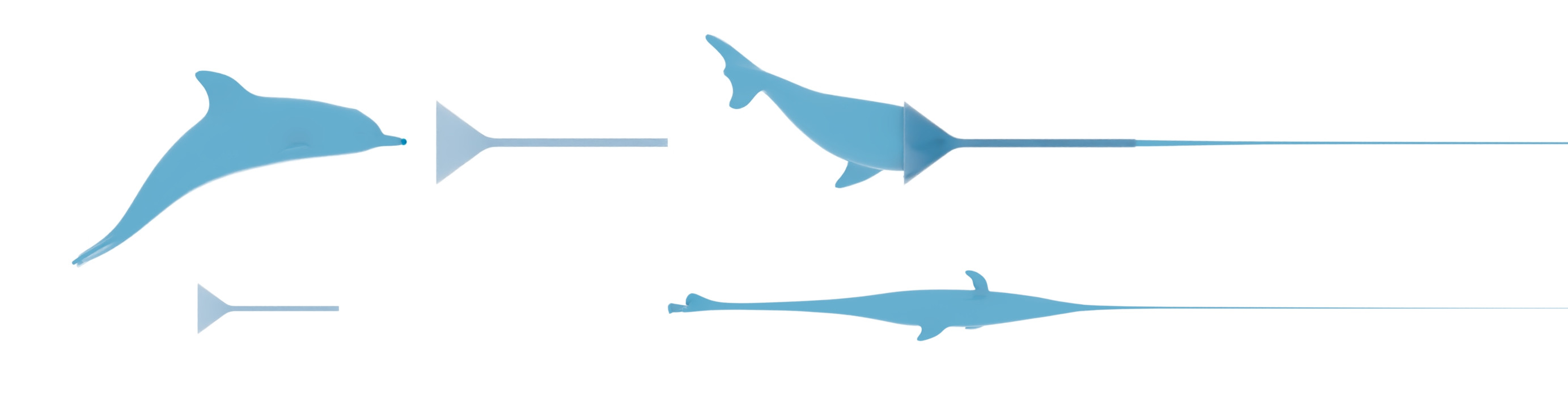}
  \caption{\textbf{Dolphin \& funnel.} The funnel test from \citet{li2020incremental}, showing a penetration-free trajectory of the dolphin passing through a narrow funnel.}
  \label{fig:funnel}
\end{figure}

\paragraph{Funnel} We also include the funnel test from \cite{li2020incremental}, in which a stiff elastic dolphin model is dragged through a long, narrow funnel (\autoref{fig:funnel}). The stiffness of the dolphin and the small funnel size make it difficult to pass through the obstacle, requiring extreme elongation to generate sufficient elastic force. Our method achieves results similar to IPC, generating a penetration-free animation of the dolphin under extreme compression and elongation.

\paragraph{Armadillo Through a Tiny Hole}  
We present another challenging example involving extreme deformation and high internal stress: an stiff Armadillo model of Young’s modulus \(E = 10\,\text{MPa}\), approximately 10 m in its longest dimension, is driven through a narrow circular hole of radius 0.24 m in a fixed plate (\autoref{fig:hole}). The motion is induced by a slender rod of radius 0.18 m acting as a moving boundary. Similar to the funnel test, this scenario requires forcing a stiff elastic body of volume \(25.7\,\text{m}^3\) through an aperture of area \(0.086\,\text{m}^2\), highlighting the robustness of our method under severe geometric and material constraints.

\subsection{Ablation Study}
\label{sec:ablation}

\begin{figure}
  \includegraphics[width=\linewidth]{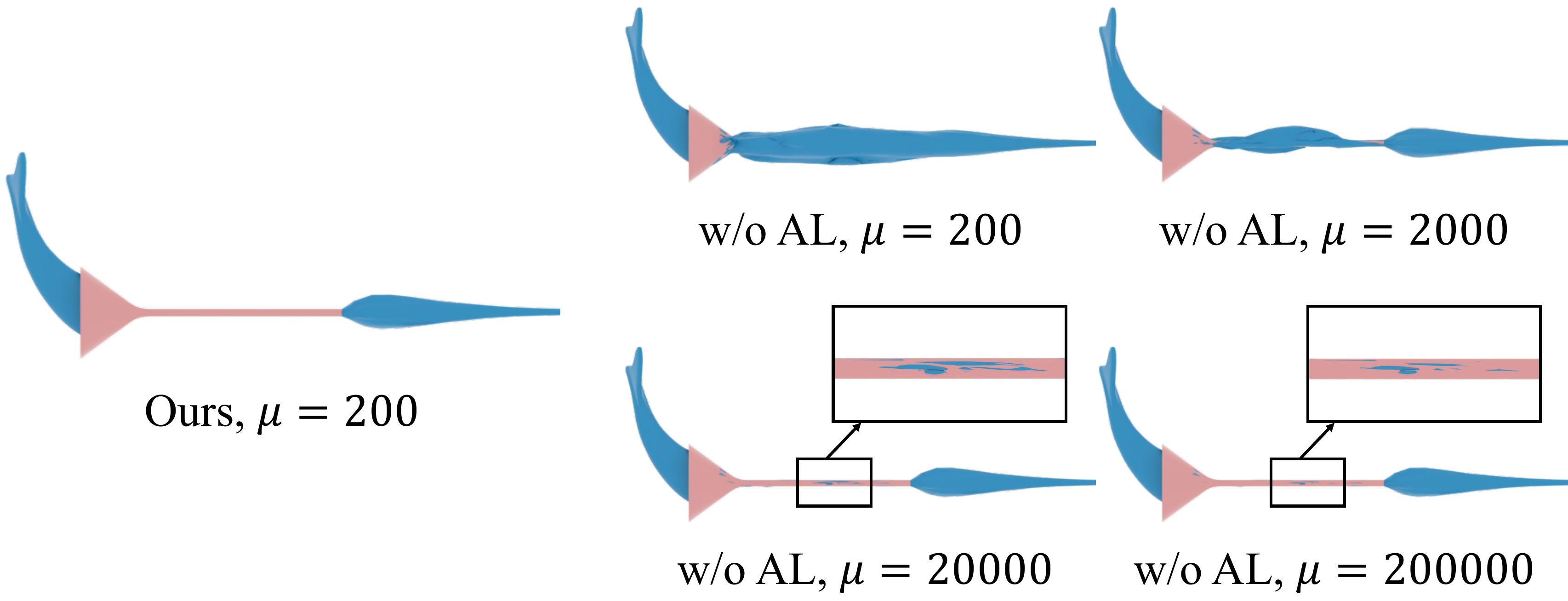}
  \caption{\textbf{Augmented Lagrangian v.s. penalty method.} Without the Lagrange multiplier term, interpenetration cannot be resolved in $\hat {\mathbf x}^{[k]}$ even with $1000\times$ larger penalty stiffness.}
  \label{fig:ablation_AL}
\end{figure}

\paragraph{Augmented Lagrangian v.s. penalty method} The efficiency and finite-step termination of our algorithm is enabled by the Augmented Lagrangian subproblem solver, which explicitly tracks the estimated Lagrange multipliers for the contact constraints and allows the use of a relatively small stiffness $\mu$ to keep the systems well-conditioned. To illustrate the effectiveness of our augmented IP formulation (\autoref{eq:total-energy}), we compare our full Augmented Lagrangian solver against using the simple quadratic penalty without the Lagrange multiplier term. Starting from a penetration-free state in the dolphin \& funnel test case, we optimize $\hat {\mathbf x}^{[k]}$ using both our full augmented IP and the simple quadratic penalty. Both methods use the same offset $\delta=1$ mm. As shown in \autoref{fig:ablation_AL}, the simple quadratic penalty generates severe penetrations in $\hat {\mathbf x}^{[k]}$ under $10\times$ larger stiffness $\mu$, and noticeable penetrations still exist even under $1000\times$ larger $\mu$. When constructing the penetration-free paths from $\{\hat {\mathbf x}^{[k]}\}_{k\ge 1}$, the penetrations in $\hat {\mathbf x}^{[k]}$ no longer decrease after the constraint set finalizes, causing the Newton iterations to stall at near-zero $\alpha^{[k]}$ and thus fail to converge.

\begin{figure}
  \includegraphics[width=\linewidth]{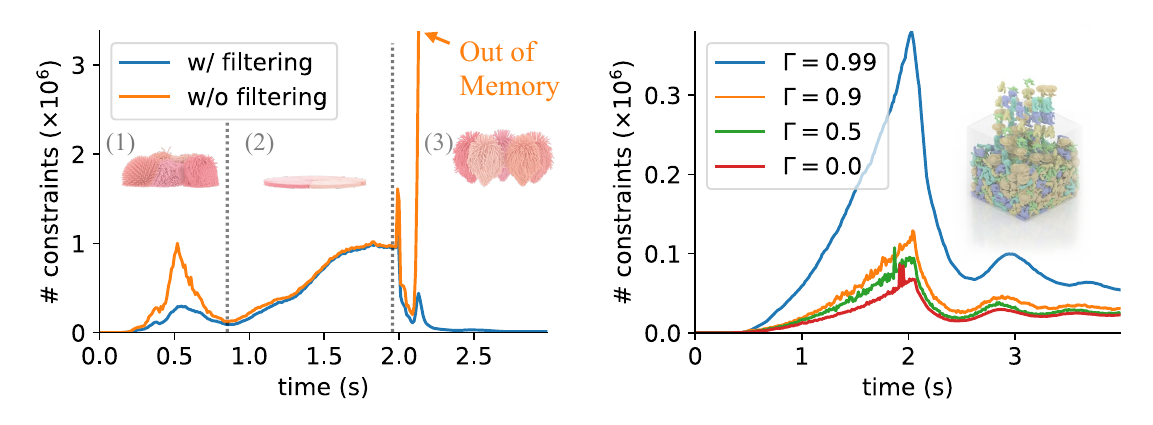}
  \caption{\textbf{Constraint filtering and decay.} Evolution of contact constraint count w/ and w/o filtering in the squishy balls example (left) and with different constraint decay factors $\Gamma$ in the animal well example (right).}
  \label{fig:decay_constraints}
\end{figure}

\paragraph{Constraint filtering} As described in Section \ref{sec:activeset}, the filtering process controls whether a newly detected contact is added to the constraint set $\mathcal C^{[k]}$. When handling contact-rich scenes or fast-moving objects, the filtering mechanism significantly reduces the number of unnecessary constraints, as shown on the left of \autoref{fig:decay_constraints}. In extreme cases (e.g., after releasing the piston in the squishy balls compression test), without filtering, the excessive number of unnecessary contacts could make the constraint set $10\times$ larger than ours, leading to higher computational cost and even out-of-memory issues.

\begin{figure}
  \includegraphics[width=\linewidth]{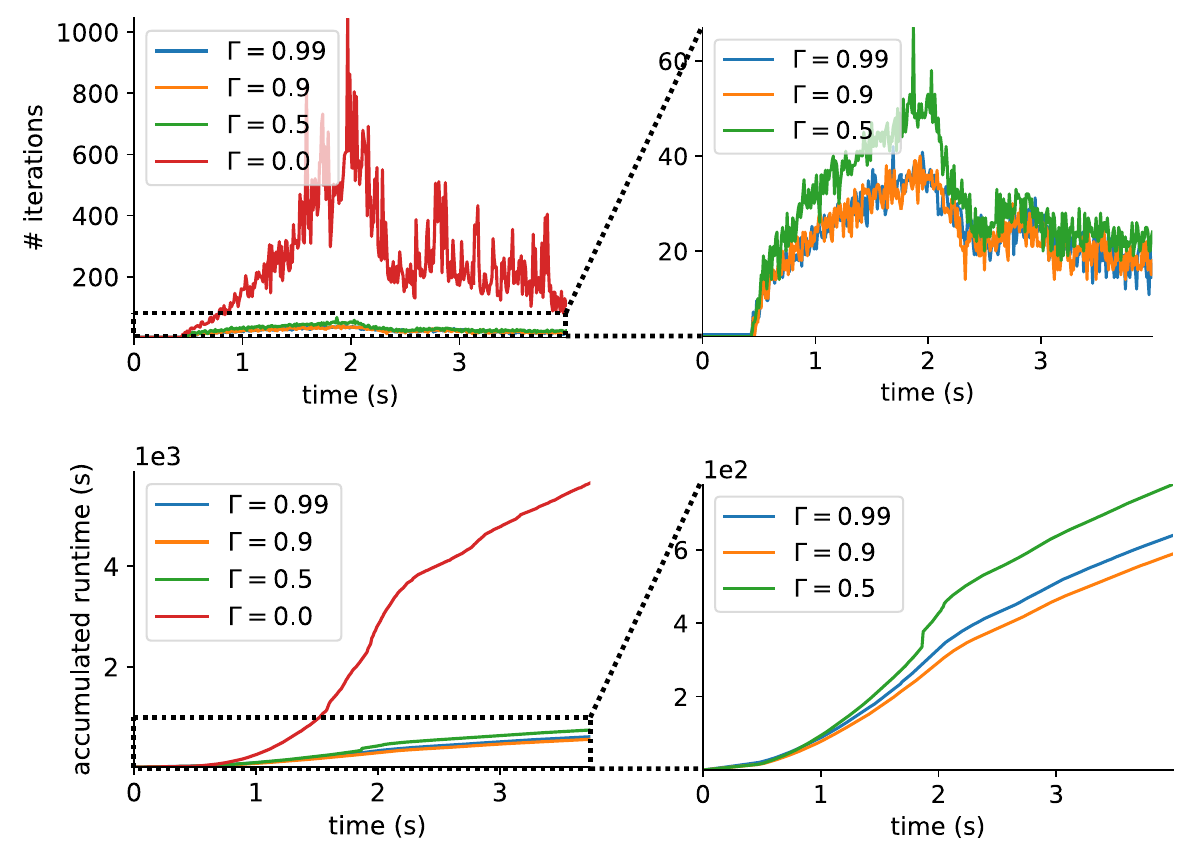}
  \caption{\textbf{Effectiveness of constraint decay.} Evolution of Newton iteration count and accumulated runtime with different constraint decay factor $\Gamma$ in the animal well example.}
  \label{fig:decay_runtime}
\end{figure}

\paragraph{Constraint decay} Just as filtering determines how $\mathcal C^{[k]}$ is properly enlarged, the constraint decay mechanism governs how constraints that become inactive are smoothly removed from $\mathcal C^{[k]}$. As stated in \S \ref{sec:decay}, we use the parameter $\Gamma$ to control the rate at which the decay factors $\gamma_i$ of inactive constraints gradually decreases, and remove them from $\mathcal C^{[k]}$ once $\gamma_i<0.01$. In the extreme case of $\Gamma = 1$, all added constraints remain permanently in $\mathcal C^{[k]}$ across time steps, thus unnecessarily wasting resources after a majority of them become inactive. \autoref{fig:decay_runtime} shows how varying values of $\Gamma<1$ affect both the number of Newton iterations and the runtime required by the subproblem solver in the animal well test. When $\Gamma = 0$, all inactive constraints are instantly removed from $\mathcal C^{[k]}$, causing the optimization objective to change drastically across iterations and potentially leading to constraints oscillating in and out of $\mathcal C^{[k]}$. This makes it significantly harder for the subproblem solver to generate large $\alpha^{[k]}$, thereby requiring far more iterations and longer runtime. The values between $\Gamma=0$ and $\Gamma=1$ show a tradeoff between the number of iterations and the constraint set size, and we select $\Gamma=0.9$ as the fixed value for our method due to its best performance. The relationship between $\Gamma$ and the size of the constraint set is illustrated on the right of \autoref{fig:decay_constraints}.

\begin{figure}
  \includegraphics[width=\linewidth]{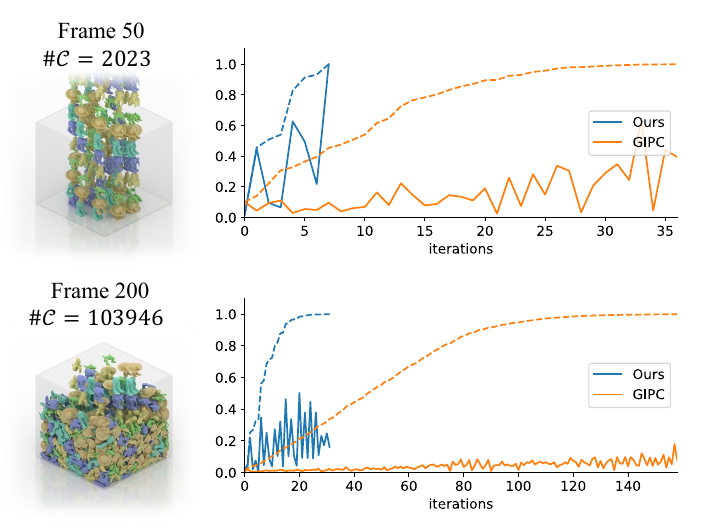}
  \caption{\textbf{Progress of termination.} Solid line: per-iteration advanced TOI $\alpha^{[k]}$. Dashed line: accumulated TOI ($1-\beta^{[k]}_0$), terminating upon reaching $1-\epsilon$. Both our method and GIPC use the same contact radius $\delta=1$ mm and termination threshold $\epsilon=10^{-3},K_\text{min}=1$. }
  \label{fig:progress}
\end{figure}

\paragraph{Progress of Termination.}
A key advantage of our method over IPC lies in the TOI progress achieved per line search, resulting in substantially fewer Newton iterations under the TOI-based termination criterion. To quantify this improvement, we first applied IPC using the TOI-based termination criterion, which reduced the number of Newton iterations per time step by an average of 2.67$\times$ in the animal well test, while maintaining comparable solution quality. To more clearly compare the progress made per Newton iteration, we compare our method and IPC (both using the TOI-based termination criterion) across scenarios with varying contact complexities. \autoref{fig:progress} plots the trajectories of $\alpha^{[k]}$ and $\beta_0^{[k]}$ until the termination condition $\beta_0^{[k]} < \epsilon = 10^{-3}$ (with $K_\text{min}=1$) is met. Unlike IPC, which truncates the search direction at the earliest detected contact and discards subsequent information, our method tracks all newly detected contacts within the penetration-allowing state $\hat{\mathbf{x}}^{[k]}$ and the constraint set $\mathcal{C}^{[k]}$. This enables significantly larger $\alpha^{[k]}$ values, particularly in contact-rich scenarios, allowing our solver to construct the penetration-free trajectory with an average of 4.24$\times$ fewer Newton iterations.

\definecolor{lightyellow}{RGB}{255, 255, 200}
\definecolor{lightred}{RGB}{255, 200, 200}

\newcommand{\yellowrowa}[8]{%
  & \cellcolor{lightyellow}#1 & \cellcolor{lightyellow}#2 & \cellcolor{lightyellow}#4 & \cellcolor{lightyellow}#5 & \cellcolor{lightyellow}#6 & \cellcolor{lightyellow}#7 & \cellcolor{lightyellow}#8}
\newcommand{\yellowrowb}[6]{%
  & \cellcolor{lightyellow}#1 & \cellcolor{lightyellow}#2 & \cellcolor{lightyellow}#3 & \cellcolor{lightyellow}#4 & \cellcolor{lightyellow}#5  & \cellcolor{lightyellow}#6\\}
\newcommand{\redrowa}[8]{%
  & \cellcolor{lightred}#1 & \cellcolor{lightred}#2 & \cellcolor{lightred}#4 & \cellcolor{lightred}#5 & \cellcolor{lightred}#6 & \cellcolor{lightred}#7 & \cellcolor{lightred}#8}
\newcommand{\redrowb}[5]{%
  & \cellcolor{lightred}#1 & \cellcolor{lightred}#2 & \cellcolor{lightred}#3 & \cellcolor{lightred}#4 & \cellcolor{lightred}#5 & \cellcolor{lightred}{-} \\}

\begin{table*}[t]
\centering
\small
\begin{tabular}{l|cc|cccccc|ccccc}
\hline
 & method & prec & \#CG & \#iters & \#contacts & TOI & time (s) & accel & Hess & PCG & CCD & LS & misc\\
\hline
\multirow{4}{*}{\autoref{fig:teaser}} 
 & Ours & double
& 28.35 & 30.09 & 0.53M & 0.192 & 5.367 & - &
0.762&1.229&3.124&0.134&0.118\\
 & GIPC & double
& 129.03 & 408.93 & 0.67M & - & 528.57 & {98.5$\times$}&
34.51&305.46&33.65&154.66&0.161\\
 & GIPC w/ TOI-term & double
& 107.58 & 137.42 & 0.62M & 0.055 & 144.22 & {26.9$\times$}&
12.72&76.59&15.07&39.78&0.062
\\
 \yellowrowa{Cubic Barrier}{single}{1}{145.24}{33.13}{1.15M}{0.163}{452.92}\yellowrowb{\textbf{$84.4\times$}}
 {414.42}{29.44}{8.605}{-}{0.453}
\hline
\multirow{3}{*}{\autoref{fig:chains}} 
 & Ours & double
& 67.10 & 16.13 & 42.6k & 0.250 & 0.657 & -&
0.025&0.114&0.499&0.007&0.012\\
 & GIPC & double & 507.33 & 261.61 & 48.3k & - & 67.77 & 103.15$\times$ &
 1.942 & 44.52& 3.432 & 18.00 & 0.017\\
 & GIPC w/ TOI-term & double & 467.56 & 167.06 & 46.3k & 0.046 & 38.88 & 59.18$\times$ &
 1.265 & 25.32 & 3.697 & 8.574 & 0.010\\
\hline
\multirow{4}{*}{\autoref{fig:animals}} 
 & Ours & double
&{53.59}&{19.23}&37.9k&{0.242}&{1.300}&-&
0.237&0.534&0.446&0.051&0.033\\
 & GIPC & double
& 144.32 & 224.75 & 47.2k & - & 115.31 & 88.7$\times$ &
15.77&69.74&9.50&20.20&0.106\\
 & GIPC w/ TOI-term & double
& 230.25 & 68.06 & 43.6k & 0.076 & 47.75&36.7$\times$&
4.881&33.78&2.951&6.048&0.090\\
 \yellowrowa{Cubic Barrier}{single}{1}{433.38}{61.04}{60.6k}{0.093}{23.30}\yellowrowb{17.9$\times$}
 {6.571}{14.56}{2.097}{-}{0.070}
\hline
\multirow{2}{*}{\autoref{fig:trapped}} 
 & Ours & double
 & 44.65 & 11.80 & 27.0k & 0.319 & 0.580&-&
 0.107&0.212&0.216&0.019&0.026\\
 & Cubic Barrier & single
 & 505.60 & 7.95 & 56.2k & 0.441 & 2.930 & 5.05$\times$ &
 0.686&2.039&0.164&-&0.041\\
\hline
\multirow{4}{*}{\shortstack{\autoref{fig:cloth_twist}\\(easy)}} 
 & Ours & double
 &24.90&2&529.67&0.833&0.0158&-&
 3.8e-3&3.2e-3&5.5e-3&1.6e-5&3.3e-3\\
 \yellowrowa{OGC ($N=10$)}{single}{10}{-}{100}{-}{-}{0.0184}\yellowrowb{1.2$\times$}{*}{*}{*}{*}{*}
 \yellowrowa{OGC ($N=50$)}{single}{50}{-}{2500}{-}{-}{0.2382}\yellowrowb{15.1$\times$}{*}{*}{*}{*}{*}
 & OGC ($N=80$)&single
 &-&6400&-&-&0.5225&33.1$\times$&*&*&*&*&*\\
\hline
\multirow{4}{*}{\shortstack{\autoref{fig:cloth_twist}\\(hard)}} 
 & Ours & double
 &51.47&2&1.8k&0.575&0.0216&-&
 3.3e-3&6.5e-3&7.3e-3&3.5e-4&4.2e-3\\
 \yellowrowa{OGC ($N=10$)}{single}{10}{-}{100}{-}{-}{0.0378}\yellowrowb{1.7$\times$}{*}{*}{*}{*}{*}
 \yellowrowa{OGC ($N=100$)}{single}{100}{-}{10k}{-}{-}{1.7618}\yellowrowb{81.6$\times$}{*}{*}{*}{*}{*}
 \yellowrowa{OGC ($N=200$)}{single}{200}{-}{40k}{-}{-}{5.8859}\yellowrowb{272.5$\times$}{*}{*}{*}{*}{*}
\hline
\multirow{4}{*}{\shortstack{\autoref{fig:cloth_stack}\\(easy)}} 
 & Ours & double
 &40.62&23.26&146.08k&0.229&1.598&-&
 0.364&0.312&0.839&0.043&0.040\\
 \yellowrowa{OGC ($N=30$)}{single}{30}{-}{900}{-}{-}{6.470}\yellowrowb{4.0$\times$}{*}{*}{*}{*}{*}
 \yellowrowa{OGC ($N=50$)}{single}{50}{-}{2500}{-}{-}{16.65}\yellowrowb{10.4$\times$}{*}{*}{*}{*}{*}
 \yellowrowa{OGC ($N=80$)}{single}{80}{-}{6400}{-}{-}{41.59}\yellowrowb{26.0$\times$}{*}{*}{*}{*}{*}
\hline
\multirow{2}{*}{\shortstack{\autoref{fig:cloth_stack}\\(hard)}} 
 & Ours & double
 & 252.71&11.84&27.73k&0.366&1.683&-&
 0.146&0.808&0.667&0.019&0.044\\
 \redrowa{OGC ($N=50$)}{single}{30\textasciitilde80}{-}{-}{-}{-}{-}\redrowb{-}{-}{-}{-}{-}
\hline
\end{tabular}\\
\raggedright
\footnotesize
\colorbox{lightyellow}{%
  \parbox[c][1pt][c]{\dimexpr\linewidth-2\fboxsep}{%
    \strut\quad \textit{Yellow:} shows noticeable artifacts or crashes near the end (see our supplementary video).%
  }%
}\\%
\vspace{-0.2mm}%
\colorbox{lightred}{%
  \parbox[c][1pt][c]{\dimexpr\linewidth-2\fboxsep}{%
    \strut\quad \textit{Red:} crashes immediately after contacts occur.%
  }%
}\\
\strut\quad * A detailed runtime breakdown of OGC is provided in the appendix.
\caption{\textbf{Comparison configurations and statistics.} TOI-term stands for TOI-based termination. OGC uses $N$ substeps and $N$ iterations per substep. {prec}: precision of floating point arithmetic. {\#iters}: average Newton (inner loop) or VBD iterations per time step. {\#contact}: average number of contact pairs. {TOI}: average advanced TOI ($\alpha^{[k]}$) for methods using TOI-based termination. {time}: average runtime per time step. {accel}: speedup factor of our method over the compared method. Hess, PCG, CCD, LS, misc: runtime breakdown of Newton-type methods into Hessian computation, linear solve, CCD, line search, and others.}
\label{tab:comparisons}
\end{table*}

\begin{table*}[t]
\centering
\footnotesize
\begin{tabular}{c|ccccccc|cccc}
\hline
 & \#V, \#F, \#T & $h$ (s) & $\delta$ (m) & model & $\rho\text{ (kg/m$^3$)},E\text{ (Pa)},\nu$ & $\mu_f,\epsilon_v$ (m/s) & $K_\text{min},\epsilon$ & \#CG & \#Newton & \#contact & time (s) \\
\hline
Fig. \ref{fig:momentum}&20.0k, 17.5k, 87.2k&0.02&1e-3&SNH&1e3, 1e6, 0.3&0.1, 1e-3&6, 1e-3
&33.6 (619)&6.0 (6)&35.8 (803)&0.16 (0.17)\\
Fig. \ref{fig:slope}&3.1k, 5.8k, 8.3k&0.01&1e-3&SNH&1e3, 1e6, 0.4&0.5$\pm$0.05, 1e-5&6, 1e-3
&30.4 (47)&6.0 (6)&434.7 (600)&0.038 (0.043)\\
Fig. \ref{fig:arch}&200, 300, 150&0.002&1e-6&SNH&1e3, 1e8, 0.3&0.5, 1e-6&6, 1e-3
&303.1 (600)&28.4 (102)&154.5 (246)&0.80 (3.52)\\
Fig. \ref{fig:fine}&49.1k, 81.9k, 160.2k&0.01&1e-2&SNH&1e3, 1e4, 0.3&-&2, 1e-3
&40.6 (50)&2.0 (2)&4.3k (8.2k)&0.15 (0.17)\\
\hline
Fig. \ref{fig:teaser}&0.87M, 1.59M, 2.25M&0.01&1e-3&COR&1e2, 1e4, 0.4&-&2, 1e-3
&28.3 (146)&30.1 (47)&0.53M (1.45M)&5.37 (12.39)\\
Fig. \ref{fig:chains}&51.0k, 87.2k, 167.4k&0.02&2e-4&COR&1e3, 1e5, 0.3&-&2, 1e-3
&67.1 (773)&16.1 (103)&42.6k (542.8k)&0.66 (56.86)\\
Fig. \ref{fig:animals}&0.43M, 0.80M, 1.34M&0.01&1e-3&SNH&1e3, 5e5, 0.3&-&2, 1e-3
&53.6 (669)&19.2 (37)&37.9k (102.6k)&1.30 (4.11)\\
Fig. \ref{fig:ramen}&37.4k, 67.6k, 95.8k&0.02&1e-3&SNH&1e2, 3.2e4, 0.3&0.1, 1e-3&6, 1e-3
&73.0 (179)&20.8 (51)&5.4k (26.6k)&0.33 (0.97)\\
Fig. \ref{fig:rods}&10.5k, 17.8k, 33.6k&0.02&2e-4&NH&1e3, 1e6, 0.3&-&2, 1e-3
&82.5 (1520)&13.4 (31)&4.7k (6.7k)&0.18 (0.45)\\
Fig. \ref{fig:rigid}&1.9k, k, 4.0k&0.02&1e-3&SNH&1e4, 1e9, 0.3&0.1, 1e-3&6, 1e-3
&200.7 (1168)&7.1 (27)&453.3 (2.7k)&0.13 (0.74)\\
Fig. \ref{fig:fall}&10.0k, 8.8k, 43.6k&0.02&1e-3&COR&1e3, 1e5, 0.49&0.5, 1e-3&6, 1e-3
&83.7 (280)&7.6 (195)&8.9k (31.3k)&0.17 (7.60)\\
Fig. \ref{fig:roller}&21.2k, 32.9k, 55.0k&0.02&1e-3&SNH&1e3, 1e7, 0.3&0.5, 1e-3&6, 1e-3
&143.0 (719)&9.4 (29)&1.2k (5.1k)&0.15 (0.45)\\
Fig. \ref{fig:hole}&21.7k, 40.2k, 64.4k&0.02&1e-3&SNH&1e2, 1e7, 0.3&-&2, 1e-3
&171.2 (3.8k)&16.9 (94)&5.8k (22.9k)&0.44 (5.83)\\
Fig. \ref{fig:funnel}&6.7k, 13.2k, 20.6k&0.02&1e-3&SNH&1e3, 1e5, 0.3&-&2, 1e-3
&71.5 (1134)&8.1 (36)&1.3k (9.4k)&0.11 (1.58)\\
\hline
Fig. \ref{fig:trapped}&0.35M, 0.64M, 0.90M&0.01&1e-3&SNH&1e3, 5e5, 0.35&-&2,1e-3&
44.7 (238)&11.8 (34)&27.0k (187.4k)&0.58 (3.57)\\
Fig. \ref{fig:cloth_twist}$^\dag$&2.5k, 4.8k, 0&1/60&2e-3&\multicolumn{2}{l}{Cloth ($\mu_\text{mem}$=1e3, $k_\text{bend}$=1e-3)}&-&2, 0.99$^*$&
24.9 (64)&2.0 (2)&529.7 (1.5k)&0.016 (0.026)\\
Fig. \ref{fig:cloth_twist}$^\ddag$&2.5k, 4.8k, 0&1/60&2e-3&\multicolumn{2}{l}{Cloth ($\mu_\text{mem}$=1e3, $k_\text{bend}$=1e-3)}&-&2, 0.99$^*$&
51.5 (217)&2.0 (2)&1.8k (4.1k)&0.022 (0.041)\\
Fig. \ref{fig:cloth_stack}$^\dag$&213.8k, 414.7k, 0&0.02&2e-3&\multicolumn{2}{l}{Cloth ($\mu_\text{mem}$=30, $k_\text{bend}$=0)}&-&6, 1e-3&
40.6 (173)&23.3 (83)&146.1k (567.5k)&1.60 (9.39)\\
Fig. \ref{fig:cloth_stack}$^\ddag$&213.8k, 414.7k, 0&0.02&2e-3&\multicolumn{2}{l}{Cloth ($\mu_\text{mem}$=300, $k_\text{bend}$=5e3)}&-&6, 1e-3&
252.7 (1670)&11.8 (43)&27.7k (772.6k)&1.68 (14.08)\\
Fig. \ref{fig:conditioning}&889, 1.3k, 3.3k&0.01&1e-3&NH&1e2, 1e5, 0.3&-&2, 1e-3&
63.1 (202)&2.1 (8)&282.0 (676)&0.028 (0.122)\\
\hline
\end{tabular}\\
\raggedright\footnotesize\quad\textit{$^\dag$ Easy version.\quad$^\ddag$ Hard version.\quad$^*$ Forced to 2 iterations.}
\caption{\textbf{Experiment configurations and statistics.} \#V, \#F, \#T: number of vertices, faces, tetrahedra. $h$: time step size. $\delta$: contact offset/radius. model: stable Neo-Hookean (SNH) \cite{smith2018stable}, Neo-Hookean (NH), corotated linear (COR) and cloth. $\rho, E, \nu$: density, Young’s modulus, and Poisson’s ratio. $\mu_f,\epsilon_v$: friction parameters. $K_\text{min},\epsilon$: termination parameters. \#CG: average (peak) CG iterations per linear solve. \#Newton: average (peak) Newton iterations per time step. \#contact: average (peak) number of contacts pairs. time: average (peak) runtime per time step.}
\label{tab:stats}
\end{table*}

\subsection{Comparisons} \label{sec:comparison}

\paragraph{GIPC \citep{huang2024gipc}} Despite benefiting from substantial GPU acceleration, GIPC, as a representation of the barrier-based IPC pipeline, remains limited by the TOI locking issue and the ill-conditioning of the barrier function. We compare our method with GIPC on two of our large-scale cases (\autoref{fig:teaser} and \autoref{fig:animals}) and one smaller case (\autoref{fig:chains}), all involving high velocities, large deformations, and challenging contacts. We test GIPC under both the original residual-based termination criterion (with default $\epsilon_d=10^{-2}l$) and the TOI-based termination criterion, using the same threshold of $\epsilon = 10^{-3}$. As shown in \autoref{tab:comparisons}, even with the TOI-based termination, GIPC requires 4.4$\times$–10.3$\times$ more Newton iterations than ours due to TOI locking. Under the same CG tolerance threshold, the ill-conditioning of the logarithmic barrier further leads to 3.79$\times$–6.96$\times$ more PCG iterations to converge. Considering all factors, including GPU optimization, our method achieves up to 103.15$\times$ faster performance than GIPC with its original residual-based termination, and 59.18$\times$ faster than GIPC with the same TOI-based termination as ours. In addition to the costs of Hessian construction and linear solves, GIPC suffers from a considerable overhead in line search as the BVH is reconstructed in every backtracking iteration. In contrast, our method naturally eliminates this expense as the constraint set is explicitly maintained and independent of the shape configuration during line searches.

\begin{wrapfigure}{l}{0.22\textwidth}
\vspace{-0.5cm}
\centering
  \includegraphics[width=0.55\linewidth]{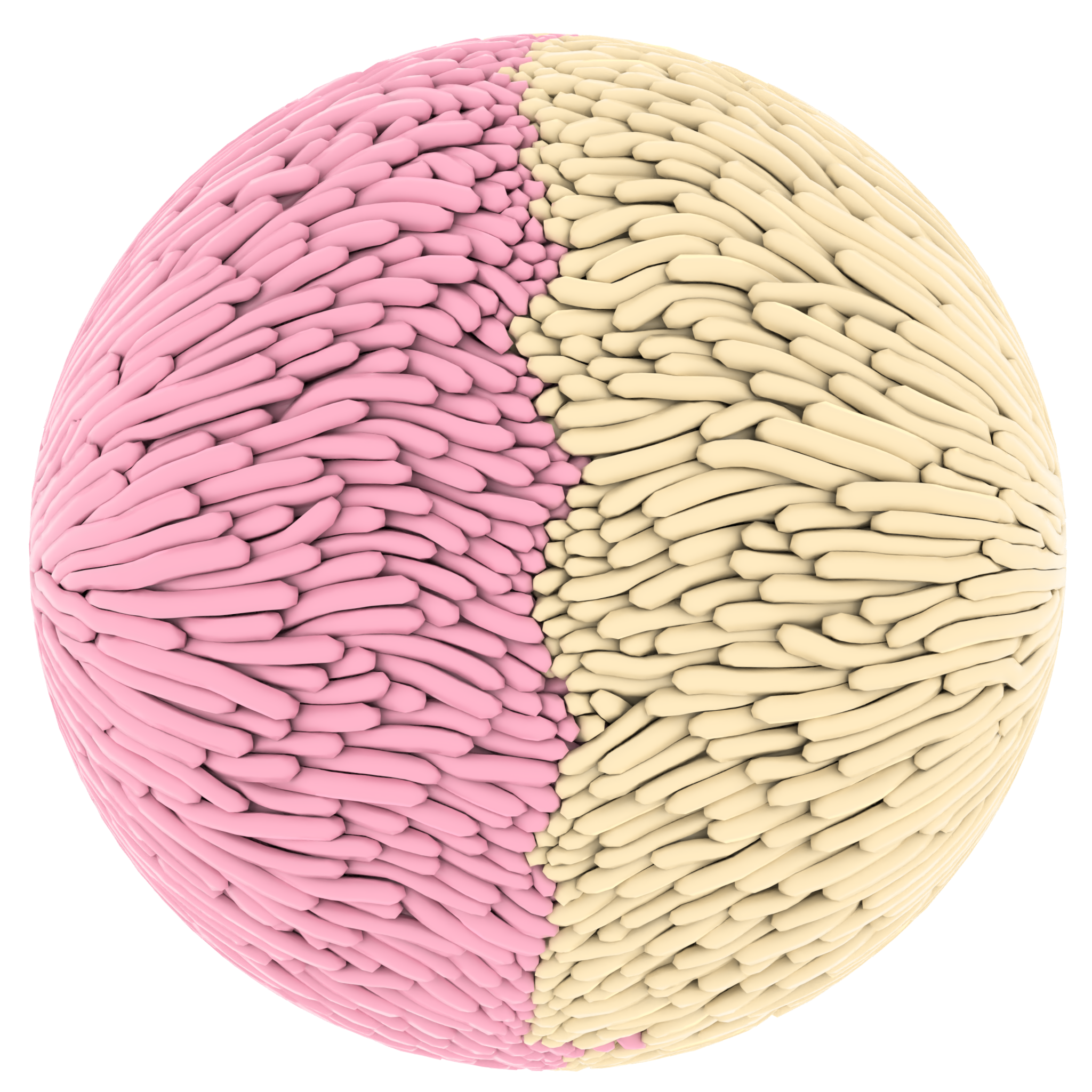}
\vspace{-0.2cm}
  \caption{\textbf{Trapped squishy balls.} Reproduced from the official repository of Cubic Barrier \cite{ando2024cubic}. }
  \label{fig:trapped}
\end{wrapfigure}

\paragraph{Cubic Barrier \citep{ando2024cubic}} In another recent work, \citet{ando2024cubic} proposed a cubic contact energy that semi-implicitly controls the contact stiffness across Newton iterations. To provide sufficient collision response, the contact stiffness of each individual contact pair is adjusted as $\kappa\gets O(d^{-2}+\rho(H))$, according to the primitive distance $d$ and an estimated spectral radius of the elasticity Hessian $H$. We compare our method with Cubic Barrier on two large-scale cases (\autoref{fig:teaser} and \autoref{fig:animals}) and a test case of trapped squishy balls (\autoref{fig:trapped}) from its official examples. For a fair comparison, we use the same TOI-based termination criterion with $\epsilon = 10^{-3}$ for both methods. Despite the performance disadvantage of double-precision arithmetic, we outperform the Cubic Barrier method (single-precision) in all cases, achieving up to 84.4$\times$ faster performance. Their design of stiffness adjustment makes the contact Hessian to scale as $O(\mu d)=O(d^{-1})$, causing the system’s conditioning diverging as the gap distance $d$ approaches zero, analogous to the IPC barrier. As shown by the statistics in Table~\ref{tab:comparisons}, this not only increases the PCG cost in each linear solve but also leads to numerical instability issues, particularly when using single-precision arithmetic: In two of our challenging cases, Cubic Barrier fails due to PCG non-convergence near the most challenging part of the simulation. We also observe that Hessian assembly becomes the dominant cost when the number of contacts is large, primarily due to the inefficient GPU implementation of CSR matrix filling. Nonetheless, we still achieve a 23.9$\times$ speedup when considering only the PCG phase in such scenarios.

\begin{figure}
  \includegraphics[width=\linewidth]{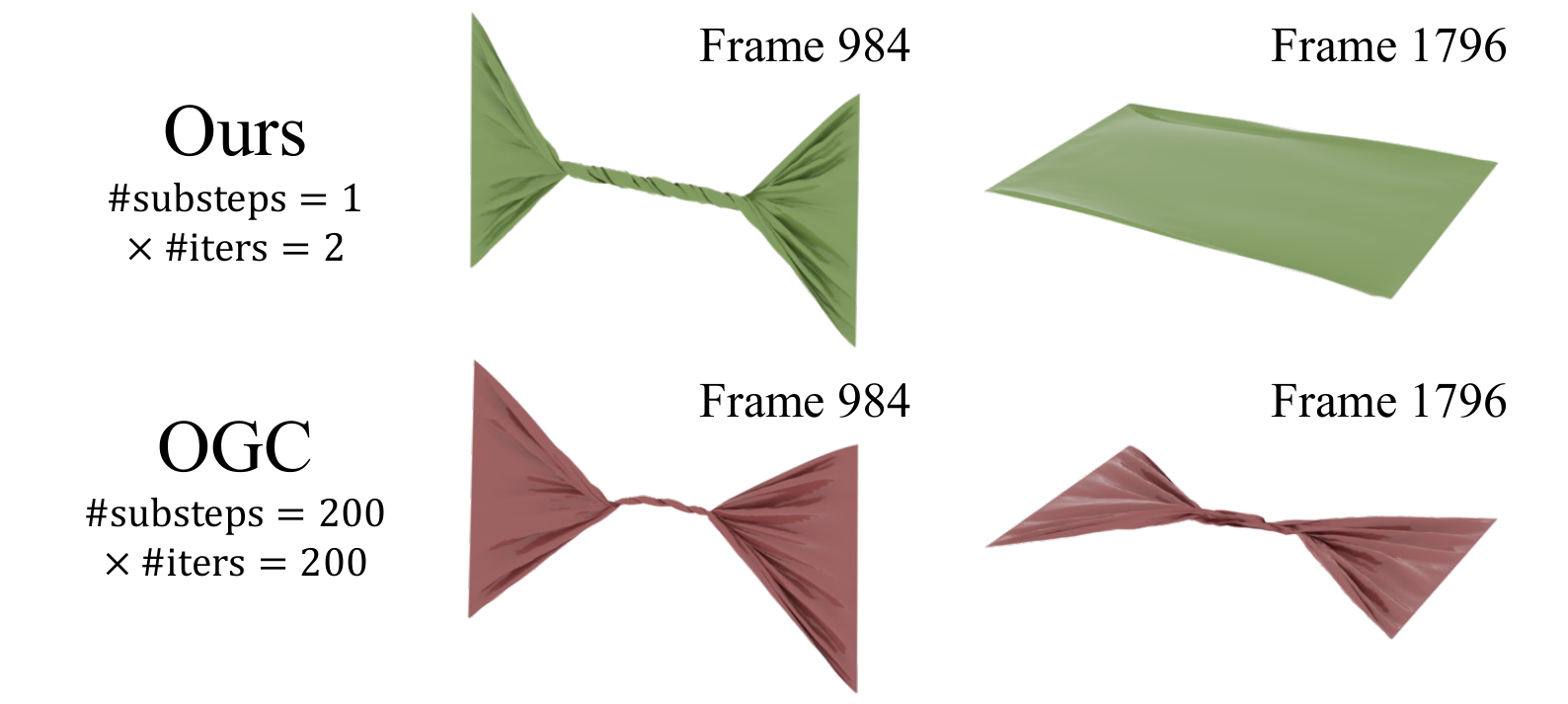}
  \caption{\textbf{Comparison to OGC -- Twisting cloth (hard version).} Both OGC \cite{chen2025offset} and our method uses the same contact radius $\delta=2$ mm. OGC fails to recover the rest shape after twisting back even with 200 substeps and 200 iterations per substep, making it 2-orders-of-magnitude slower than our method with 2 Newton iterations per time step, which produces stable and realistic results.}
  \label{fig:cloth_twist}
\end{figure}

\begin{figure}
  \includegraphics[width=\linewidth]{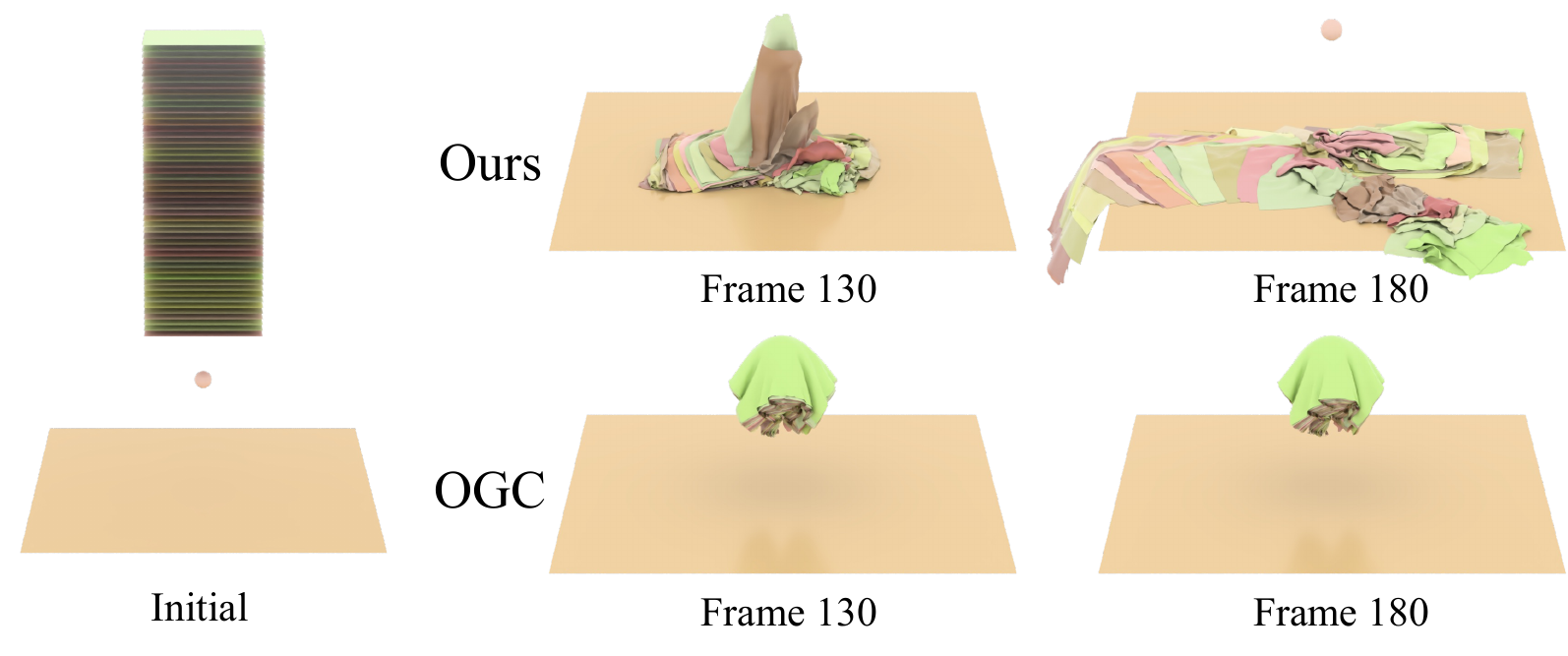}
  \caption{\textbf{Comparison to OGC -- Stacked cloth (easy version).} OGC uses $30$ substeps and $30$ iterations per time step, performing 4$\times$ slower than our method and suffering from locking artifacts with no friction. See our supplementary video for more visual comparisons.}
  \label{fig:cloth_stack}
\end{figure}

\paragraph{Offset geometric contact \citep{chen2025offset}} Offset Geometric Contact (OGC) is a recent representative work that employs first-order iterative methods on a modified IPC barrier to achieve penetration-free simulation. Despite requiring more iterations to converge, the vertex-block descent scheme used in OGC avoids solving large linear systems and thus achieves faster per-iteration performance. Instead of relying on a termination criterion, OGC provides direct control over the number of iterations per time step, offering a trade-off between speed and simulation quality. Under challenging scenarios involving intense collisions or high velocities, we show that OGC may require a very large number of substeps and iterations to achieve visually plausible results, making it less competitive than Newton-based methods with optimized linear system solvers. In addition to the sublinear convergence of coordinate descent methods, this is mainly due to the conservative bound $b_v$ used to ensure penetration-free advancement, which can become extremely small in regions with intense collisions or high mesh resolution. As a result, the motion in collision-intensive regions may be severely locked, which is especially noticeable under high velocities.

Since the official GPU implementation of OGC currently supports only co-dimensional materials, we compare it with our method on two test cases involving complex cloth self-collisions, each with two levels of difficulty. We use the parameter $N$ to control the speed–accuracy tradeoff in OGC, dividing each time step into $N$ substeps, with each substep performing $N$ iterations. The first test case of a twisting cloth (\autoref{fig:cloth_twist}) is taken directly from the examples of OGC, with all parameters kept unchanged except for the twisting duration ($6\text{ s}$ in the easy setting / $30\text{ s}$ in the hard setting) and the addition of periodic back-and-forth twisting motions. Under the easy setting, OGC fails to recover the rest shape when the twisting angle returns to zero using $N=10$ and $50$, while succeeding using $N=80$. In the harder case, OGC fails to recover the shape using all $N=10, 100, 200$. In this test, we demonstrate our method's controllability on the efficiency-accuracy trade-off while guaranteeing high-quality output. Specifically, we set our $K_\text{min}=2$ and use a large $\epsilon$ so that our method always run 2 Newton iterations per time step. As shown in \autoref{tab:comparisons}, this makes our method achieve comparable performance to OGC with $N=10$ and $33\times$ faster than OGC with $N=80$, while avoiding the mentioned artifacts.

Another test case (\autoref{fig:cloth_stack}) involves dropping $50$ layers of cloth onto a fixed sphere, either with no bending energy (easy) or with a high bending stiffness (hard). In the easy case, we outperform OGC using $N=30, 50, 80$, each suffering from different degrees of locking (see the supplementary video). In the harder case, OGC with $N=50$ crashes immediately after contacts occur, primarily due to the use of single-precision floating point and the lack of energy line search. 

\begin{figure}
  \includegraphics[width=\linewidth]{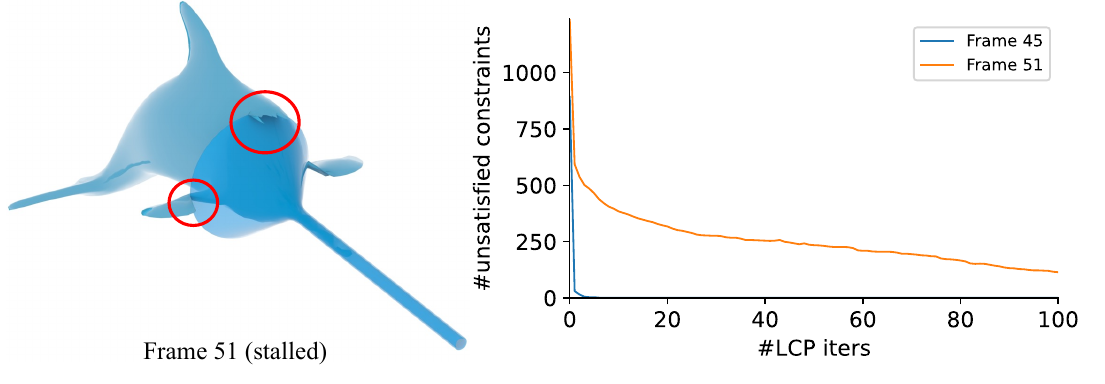}
  \caption{\textbf{Comparison to LCP-based collision handling.} Left: The LCP-based backward projection stalls at frame 51 of the dolphin \& funnel test (\autoref{fig:funnel}) and produces artifacts near the funnel's boundary. Right: Unsatisfied constraints during an LCP projection at frames 45 and 51.}
  \label{fig:lcp}
\end{figure}

\paragraph{Comparison to LCP-based collision handling} We also compare our method to an LCP-based collision handling approach similar to \citep{wang2023fast}. Specifically, we modify our main pipeline (Algorithm \autoref{alg:time-stepping}) such that, in each subproblem solve, we first advance $\hatx$ via a single Newton step using a quadratic contact penalty without the Lagrangian term. We then apply a standard projected Gauss-Seidel LCP solver to perform a mass-weighted backward projection subject to linearized non-penetration constraints. As shown in \autoref{fig:lcp}, the modified algorithm fails on the dolphin \& funnel test (\autoref{fig:funnel}), where the pipeline stalls at frame 51 with $\alpha^{[k]} \approx 0$. Although the LCP-based backward projection is capable of resolving simple contact constraints (e.g. in frame 45), its convergence is not guaranteed for more complex configurations, leading to a large portion of the constraints remaining unresolved even after 100 iterations. One possible reason is that the LCP attempts to enforce the constraints exactly, whereas the large number of linearized constraints introduced from proximity information may result in an infeasible problem. Another limitation of the LCP-based collision handling is that the mass-weighted projection is agnostic to elasticity; as illustrated in \autoref{fig:lcp}, it may produce unnatural artifacts under high stress, which are difficult to resolve in subsequent Newton iterations.

\begin{figure}
  \includegraphics[width=\linewidth]{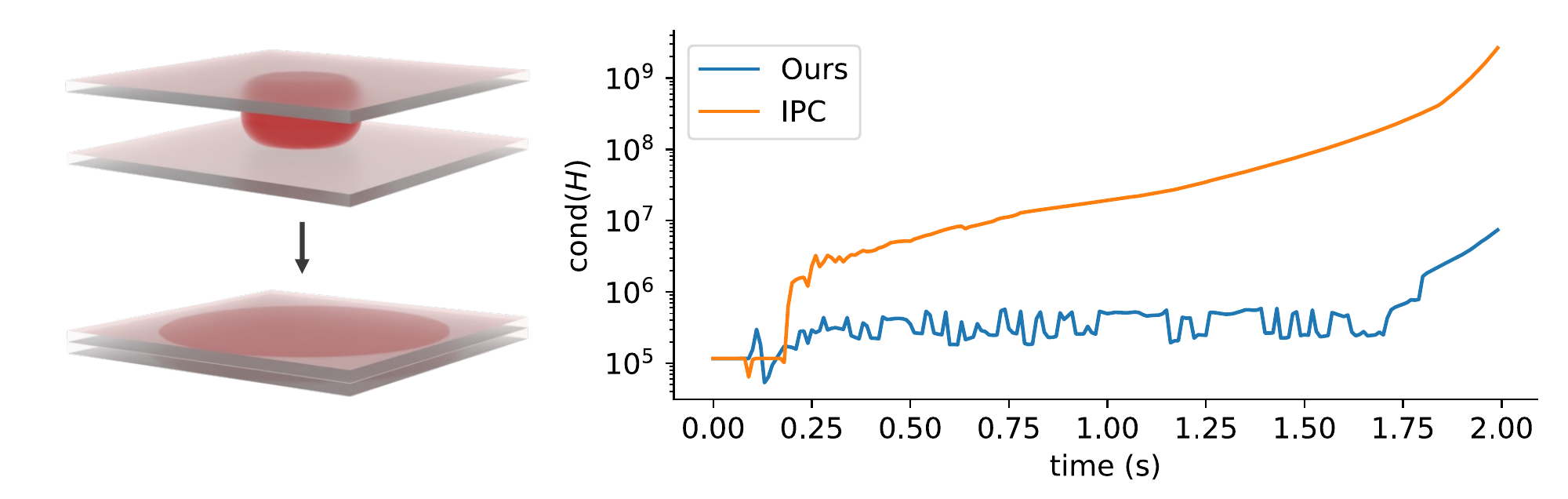}
  \caption{\textbf{System conditioning.} Unlike the IPC barrier, our contact energy remains well-conditioned under high stress until the system conditioning is dominated by the Neo-Hookean elasticity at extreme compression.}
  \label{fig:conditioning}
\end{figure}

\paragraph{Comparing system conditioning with IPC} Another advantage of our method compared to IPC is the better conditioning of linear systems, which significantly improved our performance. 
When using iterative linear solvers (e.g., PCG) in IPC, the log-barrier function becomes sharp at small distances, resulting in ill-conditioned linear systems that require specialized preconditioners \cite{huang2024stiffgipc} or a large number of CG iterations to solve. Our method avoids this conditioning issue by employing a quadratic contact penalty (\autoref{eq:total-energy}), whose Hessian for each contact pair is always congruent to a constant matrix proportional to the contact stiffness $\mu$. Taking advantage of the augmented Lagrangian method, a relatively small $\mu$ can be set to still yield good estimates of the optimal solution. 
We compare the system conditioning of our method and IPC in a simple sphere compression test, with both methods using their default estimated contact stiffness. As shown in \autoref{fig:conditioning}, our method offers a 2-orders-of-magnitude smaller system condition number in a compressing test, where even the simple block-Jacobi preconditioned CG can solve the systems in an average of 63.1 iterations.










\section{CONCLUSION}

We presented a novel barrier-free optimization framework for non-penetration elastodynamic simulation that combines an augmented Lagrangian formulation with efficient active-set exploration. 
In contrast to IPC-style approaches that rely on logarithmic barrier functions, our method models contact using a smooth augmented Lagrangian energy that remains well-conditioned under high stress and is theoretically guaranteed to provide sufficient collision response for consistent simulation progress. 
The proposed augmented Lagrangian solver, together with our novel constraint set update strategy, maintains a compact, anticipatory, and smoothly evolving constraint set, enabling rapid optimization progress without the TOI locking issues observed in IPC.
Extensive experiments demonstrate substantial performance improvements and robustness across a wide range of challenging, contact-rich scenarios. 
To the best of our knowledge, this is the fastest existing method to achieve such a high level of robustness while maintaining simulation fidelity. We believe it opens new possibilities for efficient, accurate, and reliable penetration-free simulation in time-sensitive applications such as robotics and virtual reality.

Although our method demonstrates strong efficiency and scalability, there remains significant potential for further optimization in certain components. As shown in our experiments, under collision-intensive scenarios, the cost of CCD surpasses the linear solver and becomes the dominating factor (2.69$\times$ slower than PCG in the stacked cloth test). The majority of the CCD cost arises in the broad phase during BVH queries, which currently do not exploit information from the active constraint set and therefore spend considerable time on unnecessary computations. Thus, a promising direction for further performance improvement is to design a more efficient CCD scheme optimized for our constraint set expansion. Another potential avenue for improvement lies in cloth simulation, which is currently supported by our framework but has not yet been fully optimized. In particular, developing specialized acceleration structures and constraint filtering strategies for co-dimensional geometries could further enhance both efficiency and robustness in large-scale contact-rich cloth and rod dynamics. 
In addition, applying our inequality-constrained optimization method to achieve efficient and scalable exact strain limiting would be another interesting future direction.
Finally, although we employ superlinearly convergent Newton iterations for the primal updates within our framework and achieve fast performance with high-fidelity results, the augmented Lagrangian method itself remains first-order convergent due to its dual update steps, and there is currently no theoretical guarantee on the convergence of our active set exploration strategy. Developing a more principled analysis of the active set evolution, as well as designing accelerated variants of the augmented Lagrangian solver, would be meaningful directions for future work.

\begin{acks}
We sincerely thank Kemeng Huang, Lei Lan, and Anka Chen for valuable discussion.
This work is supported in part by the Junior Faculty Startup Fund of Carnegie Mellon University and a gift from Genesis AI.
\end{acks}

\bibliographystyle{ACM-Reference-Format}
\bibliography{main}


\appendix

\section{Frictional Contact} \label{sec:friction}

Our framework supports the same semi-implicit friction model as in IPC \citep{li2020incremental}. Specifically, we include the smoothed friction potential
\begin{equation}
U_f(\x)=\sum_{i}\mu_fF_i^tf_0(\|\mathbf u_i\|;\epsilon_v)
\end{equation}
in the total potential energy $U(\x)$, where $i$ sums over all active contacts from the previous timestep, $\mu_f$ is the friction coefficient, $F_i^t$ is the normal contact force at the previous timestep, and $\mathbf u_i$ is the tangential relative displacement. The $\epsilon_v$-smoothed norm $f_0(\cdot;\epsilon_v)$ is a piecewise polynomial that satisfies $f_0(\|\mathbf u_i\|;\epsilon_v)=\|\mathbf u_i\|$ when $\|\mathbf u_i\|>h\epsilon_v$, and remains $C^2$-continuous within the region $\|\mathbf u_i\|\le h\epsilon_v$ (see \citep{li2020incremental} for details). This friction model is semi-implicit in the sense that both $F_i^t$ and the tangent operator are determined in the previous timestep and treated as constant during the current solve. We calculate the contact force via
\begin{equation}
F_i^t\gets h^{-2}\mu(c_i(\x^t)-s_i-\lambda_i/\mu),
\end{equation}
where the scaling factor $h^{-2}$ compensates for the dimensional difference between the contact penalty (\autoref{eq:total-energy}) and the potential energy $U(\x)$. Note that the decay factor $\gamma_i$ is omitted, as any constraint with $\gamma_i<1$ necessarily yields $s_i>0$, and therefore $c_i(\x^t)-s_i-\lambda_i/\mu=0$.

\begin{table*}
\centering
\small
\begin{tabular}{lrrrrr}
\toprule
& \multicolumn{2}{c}{\textbf{Twisting Cloth (Easy)}} & \multicolumn{2}{c}{\textbf{Stacked Cloth (Easy)}} & \\
\cmidrule(lr){2-3} \cmidrule(lr){4-5}
\textbf{Kernel} & \textbf{Count} & \textbf{Time (ms)} & \textbf{Count} & \textbf{Time (ms)}\\
\midrule
\texttt{memset} & 13206 & 38.888 & 1956 & 9.257 \\
\texttt{compute\_tri\_aabbs} & 161 & 0.522 & 61 & 0.720 \\
\texttt{memset\_kernel} & 4 & 0.011 & 4 & 0.012 \\
\texttt{compute\_total\_bounds} & 2 & 0.010 & 2 & 0.025 \\
\texttt{compute\_total\_inv\_edges} & 2 & 0.005 & 2 & 0.005 \\
\texttt{compute\_morton\_codes} & 2 & 0.005 & 2 & 0.012 \\
\texttt{memcpy DtoD} & 164 & 0.435 & 62 & 0.450 \\
\texttt{compute\_key\_deltas} & 2 & 0.006 & 2 & 0.008 \\
\texttt{build\_leaves} & 2 & 0.006 & 2 & 0.057 \\
\texttt{build\_hierarchy} & 2 & 0.063 & 2 & 0.530 \\
\texttt{mark\_packed\_leaf\_nodes} & 2 & 0.008 & 2 & 0.049 \\
\texttt{compute\_edge\_aabbs} & 161 & 0.503 & 61 & 1.108 \\
\texttt{apply\_rotation} & 80 & 0.386 & --- & --- \\
\texttt{bvh\_refit\_kernel} & 320 & 5.868 & 120 & 13.984 \\
\texttt{memtile\_value\_kernel} & 480 & 1.559 & 180 & 23.023 \\
\texttt{vertex\_triangle\_collision\_detection\_no\_triangle\_buffers\_kernel} & 160 & 35.468 & 60 & 164.627 \\
\texttt{edge\_colliding\_edges\_detection\_kernel} & 160 & 57.493 & 60 & 757.425 \\
\texttt{compute\_particle\_conservative\_bound} & 160 & 0.948 & 60 & 3.411 \\
\texttt{forward\_step\_penetration\_free} & 80 & 0.325 & 30 & 0.497 \\
\texttt{accumulate\_contact\_force\_and\_hessian} & 19200 & 272.613 & 4500 & 4945.037 \\
\texttt{solve\_trimesh\_with\_self\_contact\_penetration\_free\_tile} & 19200 & 175.882 & 4500 & 553.361 \\
\texttt{copy\_particle\_positions\_back} & 19200 & 66.124 & 4500 & 25.634 \\
\texttt{update\_velocity} & 80 & 0.263 & 30 & 0.224 \\
\midrule
\textbf{Total CUDA time} & \textbf{72830} & \textbf{657.400} & \textbf{16198} & \textbf{6499.458} & \\
\textbf{Total CUDA time (w/ graph)} & & \textbf{515.374} & & \textbf{6448.251} & \\
\bottomrule
\end{tabular}
\caption{CUDA kernel profiling comparison for OGC on Twisting cloth (easy, $N=80$) and Stacked cloth (easy, $N=30$).}
\label{tab:ogc_profile}
\end{table*}

\section{Comparison Details}

\subsection{Comparison Setup}

For better reproducibility of our comparative experiments, we provide the setup details of the compared methods together with additional timing breakdowns omitted from the main paper. All comparisons are conducted on a desktop PC with an Intel Core i9-13900K CPU (24 cores), 64 GB RAM, and an NVIDIA GeForce RTX 4090 GPU, the same hardware used for our method. We use the following official GPU implementations for comparison:

\begin{itemize}
\item \textit{GIPC}: \url{https://github.com/KemengHuang/GPU_IPC} (commit \texttt{405c1cc}).
\item \textit{Cubic Barrier}: \url{https://github.com/st-tech/ppf-contact-solver} (commit \texttt{919539a}).
\item \textit{OGC}: \url{https://github.com/newton-physics/newton} (commit \texttt{56c25d1}).
\end{itemize}

We made minor modifications on GIPC and Cubic Barrier to support the energy models used in our testcases. Scripts and assets to reproduce the comparison test cases are available in our supplementary materials.

\subsection{Detailed Performance Profiling for OGC}

As OGC does not provide timing breakdown by default, we employ Warp’s built-in \texttt{ScopedTimer} with full CUDA synchronization each frame to record the kernel-level timing information.

By default, OGC enables Warp's computation graph optimization to improve GPU utilization. To perform detailed kernel-level profiling, we temporarily disable the computation graph optimization, which incurs approximately a 27\% performance overhead in the smaller test case (Twisting Cloth) and a negligible overhead in the larger case (Stacked Cloth). This modification is applied solely for obtaining detailed timing breakdowns here, and the computation graph optimization remains enabled in our main comparisons (\S\ref{sec:comparison}).

We profile representative frames whose runtimes are close to the average values reported in \autoref{tab:comparisons}, using the following settings:

\begin{itemize}
\item \textbf{\autoref{fig:cloth_twist} (easy)}: Twisting Cloth with $N=80$ substeps and $N=80$ iterations per substep (a total of $6{,}400$ VBD iterations per timestep).
\item \textbf{\autoref{fig:cloth_stack} (easy)}: Stacked Cloth with $N=30$ substeps and $N=30$ iterations per substep (a total of $900$ VBD iterations per timestep).

\end{itemize}

\autoref{tab:ogc_profile} presents detailed performance profiling results for each CUDA kernel of OGC. Note that the kernels for Hessian assembly and per-vertex solving are executed $\#C\times\#\text{substeps}\times\#\text{iters}$ times, where $\#C$ denotes the number of mesh color groups.

\end{document}